# A critical cluster analysis of 44 indicators of author-level performance

Lorna Wildgaard*

*a* *Royal School of Library and Information Science, Faculty of the Humanities, Copenhagen University, Denmark, Birketinget 6, 2300 Copenhagen S*

*Corresponding author.

*Email Addresses:* pnm664@ku.dk

## Abstract

This paper explores the relationship between author-level bibliometric indicators and the researchers they "measure", exemplified across five academic seniorities and four academic domains. Using cluster methodology, the disciplinary and seniority appropriateness of author-level indicators is examined. Publication and citation data for 741 researchers across Astronomy, Environmental Science, Philosophy and Public Health was collected in Web of Science (WoS). Forty-four indicators of individual researcher performance were computed using the data. A two-step cluster analysis using IBM SPSS version 22 was performed, followed by risk analysis and ordinal logistic regression to explore cluster membership. Indicator scores were contextualized using the individual researcher's curriculum vitae. Four different clusters based on indicator scores ranked researchers as low, middle, high and extremely high performers. The results show that different indicators were appropriate in demarcating ranked performance in different disciplines. In Astronomy the *h2* indicator, *sum pp top prop* in Environmental Science, *Q2* in Philosophy and *e*-index in Public Health. The regression and odds analysis showed individual level indicator scores were primarily dependent on the number of years since the researcher's first publication registered in WoS, publications and citations. Seniority classification was secondary therefore no seniority appropriate indicators were confidently identified. Cluster methodology proved a useful method in identifying disciplinary appropriate indicators providing the preliminary data preparation was thorough but needed to be supplemented with other analyses to validate the results. A general disconnection between the performance of the researcher on their curriculum vitae and the performance of the researcher on bibliometric indicators was observed which as a result underestimated researcher performance.

# Introduction

"Quality nowadays seems to a large extent to be defined as productivity," Van Arensbergen (2014) wrote recently. Researchers are defined and are defining themselves in assessments partially in terms of their performance on bibliometric indicators of production. Developing indicators that most accurately capture the researcher's performance has led to an explosion in the amount of author-level indicators, even though it still remains the most "famous" ones like the *h-index* or *citations per paper* being used and perhaps these are not the best or most contextually appropriate ones (Wildgaard, 2015; Iliev, 2014). Several core concepts are foundational to the pursuit of the competent use of bibliometrics at the individual level. First, there is no denying that bibliometrics has become central to economic, political, social and academic evaluation systems, as well as to the individual profile of the researcher. Performance and assessment culture has been internalized and institutionalized in university and research institutions and consequently author-level indicators are being used as (self) regulatory tools to monitor and adjust scientific activities in attempts to optimize the effect of the researcher and their publications (Wouters, 2014; Retzer & Jurasinski, 2009). Although experts on bibliometric indicators do not generally see author-level indices as indicators of research quality, socially they seem to partly function like it (Van Arensbergen, 2014). Second, given the diversity of publication and citation cultures within scientific disciplines, the usefulness of an indicator is fluid (Lancho-Barrantes, 2010). An indicator that works well for one particular community of researchers is not necessarily appropriate in another community. Thus, bibliometric evaluation of the individual is framed by culturally influenced norms, disciplinary norms, and "ways of knowing" in the individual's specialty, which in turn affects the individual's visibility or coverage in generic citation databases. Third, informative bibliometric evaluation of an individual researcher requires that indicators are implemented by assessors with a high degree of understanding of the equation that is the foundation of the indicator. This improves understanding of how the indicator on a particular individual's publication/citation dataset serves as an asset or drawback in summarizing the experiences and achievements of the researcher (Sandström & Sandström, 2009). Interestingly, for many years the bibliometric community has warned about the volatility of bibliometric statistics on at the individual level as the stability of the indicators and the importance of the numbers they produce based on limited data are not stable or reliable (IEEE, 2013; Bach, 2011). Fourth, at the individual level, disciplinary and personal culture has implications for the strength or usefulness of the bibliometric indicator in evaluations—results are influenced by the age, nationality, specialty of the researcher, length of career, amount of publications, publication language, available publication and citation data, and method of data-collection. These are vital, non-consistent variables that differ from researcher to researcher and their influence must not be underestimated in useful and insightful bibliometric evaluation. But still author-level indicators are increasing in popularity.

Because of the increasing interest in author-level indicators by researchers and administrators alike, it is the responsibility of the bibliometric community to recommend useful indicators, and identify the stable indicators from the volatile and the true indicators from the spurious. The aim of this paper is to use a two-step cluster methodology to explore researcher performance measured by author-level indicators across four scientific disciplines and, if possible, recommend disciplinary and seniority appropriate indicators. The indicators used in this investigation purport to measure different aspects of research activities such as productivity, visibility, currency, impact, prestige and collaboration and should therefore accommodate different measurements of production highly relevant for researchers in all four disciplines. The research questions are these:

Which author-level indicators are most sensitive to researcher performance in different disciplines?
Which author-level indicators are most sensitive to researcher performance in different seniorities?

Further, this paper critically discusses if the applied cluster methodology actually provides an informative approach in grouping researchers and author-level metrics that we can draw conclusions from or if the results are purely arbitrary.

The cluster analysis is a process-based methodology that to give meaningful results builds on seven stages: 1) data-collection, 2) description of data, 3) calculation of bibliometric indicators, 4) presentation and statistical description of the bibliometric indicators, 5) a rationalized choice and application of the cluster algorithm and clustering statistics, 6) tests of the stability and strength of the clusters and finally, 7) informed interpretation of the clusters. These stages organize the paper as follows: The Related Literature section reviews cluster methodologies and previous bibliometric studies that have used clustering methods to study performance indicators and researcher profiles. We use the related literature to rationalize our choice of cluster methodology, stage 5. Stages 1, 2 and 3 are presented in the Methodology section, followed by stages 4 and 6 in the Results section. Finally we discuss the results, stage 7, and draw our conclusion, suggesting directions for future research on this topic.

## Related literature

Clustering and mapping techniques have similar objectives and terminology and are often used together in bibliometric analyses, in visualization of collaborations and research areas, in the development of new bibliometric software and in exploring the overlap and redundancy between indicators. However, these techniques are based on different ideas and rely on different assumptions (Waltman et al., 2010). To eliminate confusion, we begin this section by stating that the following presentation of related literature concentrates solely on cluster analysis.

Cluster analysis is concerned with exploring data and finding structure in this collection of elements that are characterized by a number of variables. The aim is to group the elements in this collection so that they are grouped in "Clusters". Each cluster contains very similar elements and is preferably highly heterogeneous from the elements grouped in the other clusters. From this clustering the internal structure of a dataset according to some chosen attributes can be interpreted. It is a tool for researchers to understand groupings of data and gain knowledge of how to classify elements in multidimensional datasets by observing their similarities and dissimilarities. The four main frameworks for cluster analysis are probabilistic, partitioning, hierarchical and hybrid clustering which are described in technical detail for example in (Ibáñez et al., 2013; Bacher et al., 2010). Within each framework there have been developed hundreds of different clustering algorithms by researchers from different scientific disciplines (Äyrämö & Kärkkäinen, 2006). Basically, each of the frameworks uses a different starting point for creating clusters. Consequently, as each of the clustering algorithms have a different composition they will produce different ordination and clustering results when used on the data and clustering results can be considered arbitrary (Schneider & Borlund, 2007b; Schneider & Borlund, 2007a). As choice of method can influence the validity of the results it is important to justify the use of the chosen method. In this paper we use the two-step clustering method and rationalize our choice by exemplifying the wide-use of clustering techniques in previous bibliometric studies and motivating the two-step method by discussing technical issues of clustering as argued in statistics literature.

The dataset we are working with consists of small collections of skewed bibliometric data, containing variables on different scales that represent individual researchers. It is a requirement of the study to be able to cluster both researchers (cases) and indicators (variables). Immediately *Probabilistic Clustering* can be eliminated as an appropriate clustering methodology. This technique only allows the clustering of cases and requires large sample sizes, e.g. n=3000 to be able to assign a case the probability of belonging to a cluster based on patterns in the data (Bacher, 1996; Bacher, 2000). Jeong and Choi (2012) explain the advantages of the probabilistic approach over partitioning and hierarchical approaches when working with bibliometric data, which include allowing uncertainty in cluster membership rather than overlapping among clusters and variability within each cluster. They found that the probabilistic analysis successful in identifying sub-components of collaboration and used this method to provide practical insights for policymakers by creating a taxonomy of taxonomy of collaboration and characteristics of type. Even though the probabilistic method supports clustering of data on different scales without the transformation of variables, the size of the dataset needed to use the probabilistic algorithm means our dataset is not large enough to support the benefits of the probabilistic approach even if it did allow the clustering of cases *and* variables.

The *partitioning* clustering method can also be eliminated. Although useful on moderately sized sample sizes (e.g. N=300), it again allows only the clustering of cases, not variables. In addition, the partitioning method typically uses the *K*-means clustering algorithm which is very sensitive to outliers and does not account for between-cluster difference leading to non-robust clusters (Äyrämö & Kärkkäinen, 2006). New methods attempt to overcome these limitations (Kaufman & Rousseeuw, 2005) and these advances together with the simplicity and computational efficiency of the partitioning method makes it an increasingly common approach in bibliometric studies, i.a. identifying research priorities in different scientific fields and partners for bilateral or multilateral cooperation in science (Chawla, 2006), and in identifying bibliographic links between journals, distance metrics and impact factors (Su et al., 2013). We thus consider *hierarchical* methods an appropriate approach for both the composition of our dataset and for the objectives of our study. Particularly the *two-step* hierarchical method allows the cluster of cases and variables, and enables the analysis of mixed scale data, for example nominal, ordinal and interval data together, which is characteristic of the indicators in our data set. Mixed scales present difficulties, and it is recommended to downgrade all data to nominal measures and use matching type coefficients to form clusters in the structure of the data (Gower, 1967). Normalizing the data in this way ensures that the chosen distance measure accords equal weight to each variable, otherwise, without normalization, the variable with the largest scale will dominate the measure. Examples of application in bibliometric studies are Otsuki & Kawamura (2013), who combine co-citation and bibliographic coupling with regional data and purchase information from Twitter to visualize purchasing behavior; and Sun et al. (2014) who combine hierarchical clustering and bibliometric analysis of integrated care literature to identify core target journals as well as create an overview of the field's key domains, indicating areas for further research and development.

Whereas the partitioning algorithm iteratively relocates data-points between clusters and updates clusters until the optimal partition is attained with no hierarchical structure (MacQueen, 1967), the two-step clustering algorithm identifies groupings by running pre-clustering first and then by hierarchical methods. The variables do not have to be on the same scale or standardized as in traditional hierarchical cluster analysis. Hierarchical clustering techniques work well with data that does not fulfill the assumption of normality and can be used on small or moderate sample sizes to produce heterogeneous clusters (Bacher, 1996; Bacher, 1996). This makes it an ideal clustering technique for exploring our dataset. Still heterogeneous clusters are perhaps not realistic when grouping bibliometric indicators. We have previously experienced first-hand how difficult it is to place indicators in just one category and rationalize for this choice (Wildgaard et al., 2014). The *hybrid* method thus proposes an alternative worth considering. In this approach each element is associated with a set of membership levels that indicate the strength of the association between that data element and a particular cluster and as a result the elements can belong to more than one cluster, and outliers and elements that link clusters together are identified (Janssens et al., 2009). This method has proved useful in classification schemes such as the Essential Science Classification that forms part of the Web of Science, where cluster analysis and cognitive mapping is integrated into subject classification (Janssens et al., 2009). Consequently the hybrid method has been suggested to produce less arbitrary clusters than the hierarchical method that attempts exclusive clustering, (Ruspini, 1970). But the hybrid method has limitations in finding the optimal cluster number and in determining why an element is placed in a cluster thus resulting in ambiguous results. We return to the two-step cluster as the suitable clustering approach. It is more resistant than the hybrid method to inter-disciplinarity and eliminates strong links with other clusters that distort the intra-cluster coherence. Unlike probabilistic clustering, it allows us to produce a hierarchical structure of clusters to identify parental relationships between bibliometric indicators on a relatively small amount of data, we have the flexibility to merge smaller clusters into larger ones (agglomerative clustering) or split larger clusters (divisive clustering) dependent on the character of our data, as exemplified in (Ibáñez et al., 2013), it accommodates variables on different scales and allows the clustering of both cases and variables, techniques not possible with partitioning approaches.

## Data collection and Methodology

*Data collection and description of data*

We collected curriculum vitae (CV), publication and citation data of 741 researchers in Web of Science (WoS) identified in an online survey conducted by Wolverhampton University in 2011 in the fields of Astronomy, Environmental Science, Philosophy and Public Health, Table 1. In total 12,359 publications and 321,443 citations in Astronomy, 7,820 publications and 118,573 citations in Environmental Science, 3,494 publications and 19,279 citations in Philosophy and 7,294 publications and 114,794 citations in Public Health. A detailed report about the data collection process and statistical description of the dataset can be found in (Wildgaard & Larsen, 2014; Wildgaard et al., 2013). In the survey the respondents reported their academic discipline and seniority, and these are used to group the researchers in our study. Additional publication and citation information on articles and reviews in this data set was kindly provided by the Centre for Science and Technology Studies (CWTS) at Leiden University, the Netherlands from their custom version of the WoS. As the CWTS data does not contain data from the Conference Proceedings Citation Indexes we do not have additional data on 3,693 citable papers and these are excluded from the present analysis. This is a unique dataset that represents different publication and citation traditions across four fields and researchers with very different publishing histories. Further is enables the critical comparisons of indicator scores and cluster placement to researcher CVs that are conducted throughout this paper.

**Table 1.** Sample of 741 researchers, distribution of publications and citations across disciplines and seniorities.

|  |  | Publications | | | Citations | | |
|---|---|---|---|---|---|---|---|
| **Discipline** | **Sample** | **Range** | **Median** | **Mean** | **Range** | **Median** | **Mean** |
| *Astrology*, 192 researchers | | | | | | | |
| *Ph.D* | 15 | 2-36 | 7 | 10.8 | 8-529 | 150 | 149.4 |
| *Post Doc* | 48 | 3-103 | 19.5 | 26 | 3-3177 | 201.5 | 561.1 |
| *Assis Prof* | 26 | 10-142 | 39.5 | 51 | 69-4009 | 702 | 1118,6 |
| *Assoc Prof* | 66 | 7-292 | 61.5 | 77.7 | 19-9083 | 1214 | 1981.1 |
| *Professor* | 37 | 34-327 | 90 | 121.3 | 177-16481 | 1889 | 3579.1 |
| *Environmental Science*, 195 researchers | | | | | | | |
| *Ph.D,* | 3 | 3-5 | 4 | 4 | 16-60 | 34 | 36 |
| *Post Doc* | 17 | 2-59 | 9 | 12.8 | 10-642 | 41 | 91.7 |
| *Assis Prof* | 39 | 2-46 | 18 | 19 | 0-573 | 148 | 185.4 |
| *Assoc Prof* | 85 | 1-103 | 29 | 36.8 | 2-2519 | 326 | 520.1 |
| *Professor* | 51 | 1-425 | 51.5 | 59.7 | 6-14141 | 435 | 998.1 |
| *Philosophy*, 222 researchers | | | | | | | |
| *Ph.D* | 8 | 1-5 | 1 | 2 | 1-33 | 0.5 | 6.2 |
| *Post Doc* | 22 | 1-31 | 4 | 7 | 0-235 | 8 | 21.4 |
| *Assis Prof* | 43 | 1-106 | 6.5 | 10.8 | 0-1829 | 6.5 | 74.3 |
| *Assoc Prof* | 74 | 1-45 | 7 | 10 | 0-565 | 8 | 50.7 |
| *Professor* | 75 | 1-140 | 18 | 28.1 | 0-3495 | 29 | 157 |
| *Public Health*, 132 researchers | | | | | | | |
| *Ph.D* | 9 | 4-27 | 8 | 12.2 | 7-253 | 60 | 82.2 |
| *Post Doc* | 14 | 1-23 | 11 | 12 | 0-353 | 80.5 | 113.6 |
| *Assis Prof* | 30 | 3-288 | 22 | 36.2 | 10-3796 | 167 | 417.4 |
| *Assoc Prof* | 50 | 4-221 | 43 | 54.6 | 4-3649 | 518 | 778.5 |
| *Professor* | 29 | 5-661 | 76 | 110.2 | 13-13520 | 954 | 2104 |

*Method*
IBM SPSS version 22 was used for calculation of statistics. We performed a statistical description of each indicator to explore the range and spread of indicator scores and consequently interpret how indicator scores summarize the performance of the researchers. A two-step cluster analysis followed to segment the researchers, resulting in a four cluster solution that divided the researchers into four groups: extremely high (the outliers), high, middle, low scores on the bibliometric indicators. The model fit was fair to good across all fields. We used *F*-test statistics to investigate the importance of each indicator as a predictor of a researcher being placed in a cluster and the mean values of each indicator to summarize similarities and dissimilarities between clusters within each field. Odds ratios were calculated to analyse the likelihood of a researcher of a specific academic seniority being placed in a specific cluster and likewise an ordinal logistic regression was performed to assess the impact of academic age and seniority in cluster placement. This final exploration was conducted because we are concerned with how coverage or indexing practices in the database used to source publication and citation data, can distort interpretation of researcher prestige. Finally, we performed correlations of researchers measured on complementary indicators to visualize the difference between and within clusters and also to observe changes in researcher rankings.

*Definition of bibliometric indicators*

A full description of the 44 bibliometric indicators included in this analysis are presented in Appendix 1, and are briefly summarized below. We have previously discussed these indicators in detail in (Wildgaard et al., 2014; Wildgaard, 2015):

1) *Publication-based indicators* indicate the productivity of the researcher *P* and *Fp*. While *App, mean pp collab* and *mean pp int collab* indicate the extent of collaboration by extracting information from the author bylines of the analyzed articles.

2) *Citation-based indicators:*
    a. *Citation count:* indicate the visibility or effect of the researcher's publications within their academic specialty. Effect is counted as citations, as in *C*, *CPP*, *Csc, sc, nnc, SIG,* and *Cless5*. *AWCR, Cage* and *PI* are adjusted for the age of the publications, while *Fc, FracCPP* and *AWCRpa* normalize for the number of authors written on the author byline of each paper.
    b. *Citation count normalized to publications and field:* Indicators that compare the researcher's citation count to expected performance in their chosen field, *sum pp top n cits, sum pp top prop*, *NprodP* and *T>ca*.
    c. *Effect of output as citations normalized to publications and portfolio:* Indicators that normalize citations to the researcher's portfolio, *%sc* and *%nc*.

3) *Indicators that indicate prestige using Journal Impact measures*: impact of a researcher's chosen journals to suggest the potential visibility of the researcher's work in the field in which he/she is active, *mcs, mncs, mean mjs mcs, max mjs mcs* and *mean mjs*. Journal categories in the citation index, are used as the proxy for scientific fields.

4) *Hybrid indicators:* indicators of the level and performance of all of the researcher's publications or selected top performing publications. These indicators rank publications by the amount of citations each publication has received and establish a mathematical cut-off point for what is included or excluded in the ranking. They can be subdivided into indicators that are:
    a) *dependent on the calculation of the h index*: $h$, $\hbar$, $Q2$, $h2,m,A,$ the *e*-index which supplements the *h*-index by computing the value of highly cited papers while *hg* allows a greater granularity in comparison between researchers with similar *h*- and *g*- indicators.
    b) *h "independent" indicators*: *AW* is the age-weighted indicator suggested for comparison with the *h*-index and the *g*-index allows greater distinction between the order researchers.
    c) *h adjusted to field: hnorm* allows across field comparison for multidisciplinary researchers,
    d) *h adjusted for co-authorship*: POPh, and,
    e) *h-type indicators of impact over time:* indicators of the extent a researcher's output continues to be used or the decline in use, *AR index, M*-quotient and *mg*-quotient are respectively the *h* and *g* indices divided by the academic age of the researcher.

# Results

*Statistical description of indicators*

Preliminary data exploration revealed disciplinary publication trends that influence coverage in WoS and accordingly indicator scores: *Astronomy* has a strong preference for multi-authorship in article and conference publications, 12,359 out of 21,109 total publications reported on the researcher CVs were identified in WoS (58%); *Environmental Science* also publishes a great amount of article, conference papers and EU project reports and 47% of the researchers publications were included in WoS (7,820 /16,720). *Philosophy* is a dialogue-based discipline, preferring single authorship publishing in blogs, in the media, books and in national languages, coverage in WoS 3,494 out of 14,724 publications listed on CVs, (24%). *Public Health* has a strong tradition of publishing articles in international journals in collaboration with medical researchers but also publishes a lot of articles and reports in local journals in national languages on local health issues and regulations, 7,294 out of 9,067 publications were identified, (80%).

We computed all bibliometric indices for all researchers in all four disciplines, Appendix 2. Astronomy researchers typically had the highest indicator scores, Public Health and Environmental Science the next highest and Philosophy the lowest. The range and distribution of scores was similar across disciplines – the majority of researchers scores lie in the $50^{th}$ percentile, proportionally distributed around the median. The median values for *mncs*, *mnjs*, the *Price Index* and the *NprodP* indicator were the same across all four disciplines, however the variability of the scores within each group were vastly different. Philosophy had the highest percent of non-cited publications (*%nc*). The pattern of score distribution was very similar across the remaining indicators, and consequently we exemplify the range of scores across and within the disciplines using just one indicator, the most famous indicator of individual performance, *h*. A score of *h*1 was the lower whisker in Astronomy and *h*0 in Environmental Science, Philosophy and Public Health. The values of the upper whisker were *h*49.8, *h*27.5, *h*8.5 and *h*32.5 for the four disciplines respectively. The $25^{th}$ percentile, $50^{th}$ percentile and $75^{th}$ percentile were *h*8.0, *h*15.0 and *h*24.7 for researchers in Astronomy; *h* 5.0, *h* 9.0 and *h* 14.0 for Environmental Scientists; *h*1.0, *h* 2.0 and *h* 4.0 for Philosophers and, *h* 5.0, *h* 9.0 and *h*16.0 for researchers in Public Health. In bibliometric studies outliers are typically the interesting cases. Outliers on *h* were identified, and the bibliometric data compared to the "outlier" researcher's CV. In Astronomy the outlier was a highly awarded British Professor with over 650 publications listed on the CV and a career span of 34 years. In our dataset extracted from WoS this professor had 279 papers that in total attract 16,481 citations, the most significant paper attracting 1,365 citations, resulting in *h*66. This researcher's first publication in WoS was from 1980, giving an academic age of 33 years (year of data collection 2013 minus 1980). In Environmental Science, the outlier was a Dutch Professor (*h*59), specializing in catalysis engineering with involvement in many EU projects and patents, a strong presence on the web and 466 publications listed on the CV with a career of 39 years. In the WoS dataset this scientist had an academic age of 32 years, 425 papers attracting 14,141 citations, the most significant paper scoring 503 citations. While in Philosophy the outlier is a British Professor of the philosophy of economics, with a career identified from the CV of 39 years and over 225 publications. In WoS the academic age was 30 years, with 113 papers attracting 1279 citations, *h*20, the most significant paper attracting 183 citations. Finally in Public Health a Dutch Professor specializing in Medical Statistics was the outlier with 321 publications on the CV and a career span of 27 years. This researcher's academic age in the WoS dataset was 21 years, 187 papers were identified attracting 13,520 citations, *h*60, the most significant paper attracting 447 citations.

*Two-step cluster analysis of indicators*

A four cluster solution gave a fair model of the data in all fields. Table 2 shows all researchers grouped in the four clusters, the number of researchers in each cluster (size) and the average academic age of the researchers in each group. Academic age is calculated as the number of years since the researcher's first publication registered in the WoS. The range of academic ages is shown in parenthesis.

**Table 2**. Segmentation of researchers, 4 cluster solution.

| Cluster | Description | Astronomy | | Environmental Sci | | Philosophy | | Public Health | |
|---|---|---|---|---|---|---|---|---|---|
| | | Size | Academic Age | Size | Academic Age | Size | Academic Age | Size | Academic Age |
| 1 | Below interquartile range | 53 | 7 (2;18) | 63 | 9(2;31) | 132* | 9 (1;33) | 59 | 9 (1;25) |
| 2 | Median interquartile range | 58 | 21 (10:33) | 86 | 17 (7;36) | 72 | 15 (3;33) | 48 | 15 (4;34) |
| 3 | Top of interquartile range | 62 | 15 (3-34) | 45 | 22 (9;34) | 14 | 18 (7-33) | 21 | 21 (10;33) |
| 4 | Extreme outliers | 19 | 23 (11-33) | 1 | 32 | 4 | 22 (15;30) | 4 | 22 (19;28) |

*low values were below median but within interquartile range

Boxplot visualization of the indicators within each cluster showed that the indicator scores for researchers grouped in Cluster 1 were low scores below the interquartile range, in Cluster 2 between the median and the upper line of the boxplot, Cluster 3 included researchers whose indicator values lay a little higher than the interquartile range and finally Cluster 4 grouped the researchers with extremely high scores. The distribution of bibliometric data used to compute each indicator within the cluster was skewed to the left. For all indicators and across all disciplines, the researchers with the lowest scores (Cluster 1) were in the tail end of the skew to the left, the median scores (Cluster 2) were more evenly distributed slightly left of center, the high scores (Cluster 3) approached a normal distribution and the extreme scores (Cluster 4) lay in the right tail of the distribution.

The importance of each indicator as a predictor of the cluster was investigated, i.e. which indicator distinguishes the cluster. Based on the *F*-test statistic, scores range between 0-1, the closer to 1 the less likely the variation for a variable between clusters is due to chance and more likely due to some underlying difference (IBM, 2012). The post hoc test Tamhane T2 was applied to compute *F*-test statistic where equal variance in the data is not assumed and sample sizes are unequal. The following indicators were identified with predictor importance =1: In Astronomy scores on the *h2-index* determined cluster membership, in Environmental Science it was the *sum pp top prop,* in Philosophy *Q2* and in Public Health the *e*-index. Table 3 presents indicators which mean indicator values statistically discriminate between Clusters 1, 2, 3 and 4, ranked in order of importance for researcher inclusion in the cluster (001 alpha level):

**Table 3**. Indicators with statistically different mean values*

| Field | Indicator |
|---|---|
| Astronomy | *h2, ℏ, g, hg, e, AR, Q2, h, AW, Cless 5, A, sc, m, C, AWCR, sum pp top n cits.* |
| Environmental Science | *sum pp top prop, C, Csc, AWCR, sum pp top n cits, g, ℏ, Q2, hg, h2, h, AW, e, POPh, AR, sc, A, AWCRpa, P, Fc, FracCPP, CPP, Fp, mcs, mean mjs mcs, mg, max mjs mcs, SIG, cage, PR, %nc* |
| Philosophy | *Q2, g, ℏ ,hg, e, Fc, AW, AWCRpa, h, Cless5, h2, C, AWCR, Csc, AR, sum pp top prop* |
| Public Health | *e, A, g, h2, m, ℏ, Q2, AR, hg, Sig, AW, h, Csc, poph, C, Fc, AWCRpa, Cless5, AWCR, sum pp top n cits, sc, P, sum pp top prop,* |

*Note: Philosophy and Public Health only showed statistic difference between Clusters 1, 2 and 3. As Environmental Science had only 1 researcher in Cluster 4, this cluster was not included in the analysis.

**Table 4.** Mean indicator scores within the four Clusters (1=low, 2=median, 3=top, 4=extreme) per field. Indicators used in projection figures are marked in bold

| | Astronomy | | | | Environmental Sci. | | | | Philosophy | | | | Pub. Health | | | |
|---|---|---|---|---|---|---|---|---|---|---|---|---|---|---|---|---|
| Cluster | 1 | 2 | 3 | 4 | 1 | 2 | 3 | 4 | 1 | 2 | 3 | 4 | 1 | 2 | 3 | 4 |
| P | 15.9 | 53.4 | 70.4 | 213.2 | 10.6 | 37.6 | 77.3 | 425.0 | 6.4 | 24.1 | 35.9 | 99.7 | 17.3 | 48.2 | 140.7 | 249.7 |
| **Fp** | **5.9** | **26.8** | **24.6** | **60.1** | **5.2** | **16.5** | **34.7** | **136.2** | **6.2** | **22.1** | **25.2** | **61.9** | **7.6** | **17.1** | **41.9** | **68.9** |
| App | 4.1 | 3.6 | 5.4 | 6.6 | 2.9 | 3.3 | 3.5 | 3.9 | 1.1 | 1.5 | 2.2 | 2.6 | 3.5 | 4.4 | 5.0 | 4.6 |
| Mean pp collab | 0.7 | 0.7 | 0.8 | 0.8 | 0.5 | 0.5 | 0.5 | 0.4 | 0.1 | 0.2 | 0.5 | 0.5 | 0.6 | 0.6 | 0.7 | 0.6 |
| Mean pp int collab | 0.7 | 0.7 | 0.7 | 0.8 | 0.2 | 0.3 | 0.3 | 0.3 | 0.0 | 0.1 | 0.3 | 0.4 | 0.3 | 0.3 | 0.4 | 0.3 |
| **C** | **150.9** | **741.7** | **1990.7** | **7736** | **46.6** | **361.4** | **1564.6** | **14141.0** | **5.6** | **65.5** | **423.7** | **1971.2** | **97.5** | **591.3** | **2602.9** | **6499.2** |
| Csc | 88.7 | 484.9 | 1357.6 | 5133.1 | 33.8 | 259.9 | 1237.5 | 11750.0 | 4.1 | 53.1 | 331.3 | 1528.2 | 72.1 | 468.9 | 2077.7 | 5593.2 |
| Sc | 62.2 | 256.8 | 633.1 | 2.603 | 12.7 | 101.5 | 327.0 | 2391.0 | 1.4 | 12.4 | 92.3 | 443.0 | 25.3 | 122.3 | 525.1 | 906.0 |
| %sc | 40.6 | 34.8 | 33 | 35.7 | 27.0 | 27.7 | 21.0 | 16.9 | 17.8 | 20.4 | 22.1 | 22.9 | 25.4 | 19.4 | 20.0 | 11.7 |
| Nnc | 2.1 | 5.6 | 4.7 | 14.3 | 2.6 | 5.6 | 7.6 | 33.0 | 4.1 | 12.8 | 8.7 | 18.2 | 6.0 | 12.1 | 30.9 | 69.7 |
| %nnc | 14.7 | 10.9 | 6 | 6.2 | 24.9 | 14.8 | 9.4 | 7.7 | 64.7 | 41.4 | 24.9 | 18.1 | 37.7 | 24.6 | 20.6 | 29.7 |
| **CPP** | **9.0** | **14.0** | **30.2** | **36.6** | **4.7** | **10.2** | **21.6** | **33.2** | **0.8** | **3.5** | **14.2** | **19.7** | **5.2** | **14.1** | **20.1** | **45.6** |
| Sig | 34.2 | 98.8 | 231.0 | 566.0 | 15.7 | 57.9 | 198.6 | 503.0 | 2.9 | 19.7 | 140.0 | 220.7 | 26.8 | 105.3 | 269.8 | 810.2 |
| Cless5 | 122.0 | 453.9 | 1325.1 | 5013.8 | 304.7 | 316.6 | 439.5 | 21.0 | 3.8 | 38.2 | 230.5 | 1044.5 | 74.6 | 394.8 | 1500.2 | 3743.5 |
| PI | 80.9 | 62.3 | 69.5 | 66.6 | 79.1 | 65.6 | 55.2 | 49.2 | 55.0 | 65.5 | 57.6 | 54.1 | 79.9 | 67.5 | 58.3 | 49.5 |
| Cage | 1.6 | 3.7 | 3.0 | 3.4 | 2.0 | 3.1 | 4.2 | 4.9 | 1.2 | 2.1 | 3.1 | 3.5 | 1.4 | 2.4 | 3.2 | 3.0 |
| Fc | 49.9 | 326.7 | 664.7 | 2089.8 | 21.9 | 150.7 | 707.4 | 4494.5 | 5.3 | 54.3 | 211.0 | 1230.2 | 37.7 | 192.9 | 753.0 | 1514.8 |
| FracCPP | 6.8 | 11.2 | 22.4 | 28.6 | 3.5 | 7.4 | 17.2 | 27.4 | 0.4 | 4.2 | 16.7 | 14.5 | 6.3 | 10.2 | 13.7 | 33.7 |
| Sum pp top n cits | 2.8 | 15.2 | 31.7 | 102.0 | 0.9 | 7.5 | 32.9 | 235.0 | 0.0 | 1.4 | 7.8 | 36.0 | 2.2 | 13.4 | 52.8 | 89.2 |
| **H** | **6.3** | **14.7** | **22.9** | **44.8** | **3.6** | **10.1** | **20.8** | **59.0** | **1.1** | **4.1** | **9.4** | **23.5** | **4.9** | **12.3** | **26.6** | **32.2** |
| AWCR | 35.7 | 86.6 | 286.0 | 993.1 | 9.1 | 50.28 | 162.4 | 1425.1 | 0.8 | 8.2 | 47.0 | 204.7 | 19.1 | 84.4 | 299.5 | 781.4 |
| AWCRpa | 11.7 | 35.9 | 78.1 | 212.9 | 4.2 | 21.5 | 66.9 | 400.8 | 0.8 | 7.1 | 33.1 | 111.7 | 8.5 | 30.1 | 90.3 | 211.8 |
| G | 10.2 | 23.5 | 40.3 | 77.2 | 5.4 | 16.3 | 35.4 | 99.0 | 1.1 | 6.5 | 18.8 | 40.5 | 8.0 | 21.7 | 45.2 | 68.7 |
| H2 | 4.8 | 8.6 | 12.1 | 19.3 | 3.3 | 6.8 | 11.3 | 24.1 | 1.3 | 3.7 | 7.3 | 12.2 | 4.1 | 8.1 | 13.6 | 17.5 |
| POPh | 2.5 | 6.7 | 9.2 | 14.8 | 1.7 | 4.7 | 9.9 | 27.0 | 0.6 | 2.7 | 5.2 | 13.0 | 2.3 | 5.4 | 10.1 | 13.2 |
| **M-quot** | **0.9** | **0.7** | **1.7** | **2.0** | **0.5** | **0.6** | **0.9** | **1.8** | **0.1** | **0.3** | **0.6** | **1.1** | **0.6** | **0.9** | **1.3** | **1.5** |
| Mg-quot | 1.4 | 1.2 | 3.0 | 3.5 | 0.7 | 1.0 | 1.7 | 3.0 | 0.1 | 0.5 | 1.3 | 1.8 | 1.0 | 1.7 | 2.3 | 3.3 |
| Q2 | 9.2 | 19.3 | 32.2 | 60.4 | 5.2 | 13.7 | 28.0 | 82.3 | 1.6 | 6.0 | 15.1 | 31.9 | 7.3 | 17.8 | 35.4 | 51.6 |
| AW | 5.3 | 8.8 | 16.1 | 30.8 | 2.8 | 6.8 | 12.4 | 37.7 | 0.7 | 2.7 | 6.6 | 13.8 | 3.9 | 8.8 | 17.0 | 25.0 |
| E | 7.6 | 16.0 | 29.7 | 52.6 | 4.2 | 11.4 | 25.1 | 59.6 | 0.9 | 4.9 | 15.3 | 28.7 | 6.1 | 16.4 | 31.7 | 52.6 |
| **M** | **14.0** | **25.7** | **46.1** | **81.9** | **8.3** | **19.1** | **38.2** | **115.0** | **2.5** | **9.3** | **29.0** | **43.5** | **11.2** | **27.2** | **47.5** | **96.1** |
| A | 16.4 | 33.0 | 64.0 | 117.6 | 9.4 | 24.2 | 53.1 | 149.7 | 2.6 | 10.9 | 41.2 | 59.4 | 13.3 | 37.0 | 66.0 | 160.4 |
| AR | 3.9 | 5.6 | 7.9 | 10.6 | 2.9 | 4.8 | 7.2 | 12.2 | 1.3 | 3.2 | 6.2 | 7.6 | 3.5 | 6.0 | 8.1 | 12.5 |
| Hg | 8.0 | 18.6 | 30.3 | 58.7 | 4.4 | 12.8 | 27.1 | 76.4 | 1.0 | 5.1 | 13.1 | 30.8 | 6.2 | 16.3 | 34.6 | 46.3 |
| **hnorm** | **1.2** | **0.5** | **0.8** | **0.3** | **0.7** | **0.5** | **0.5** | **0.1** | **0.2** | **0.4** | **0.5** | **0.3** | **0.5** | **0.5** | **0.2** | **0.3** |
| ℏ | 7.9 | 18.4 | 30.2 | 60.7 | 4.4 | 12.9 | 27.9 | 84,0 | 1.3 | 5.3 | 14.2 | 30.6 | 6.2 | 16.6 | 35.6 | 53.4 |
| Sum pp top prop | 0.9 | 2.4 | 10.8 | 36.7 | 0.3 | 2.6 | 10.1 | 115.4 | 0.1 | 1.9 | 4.5 | 15.5 | 1.0 | 4.7 | 16.1 | 35.6 |
| NprodP | 8.8 | 44.1 | 36.1 | 129.8 | 14.1 | 47.45 | 72.2 | 580.6 | 91.4 | 220.1 | 60.3 | 82.5 | 17.5 | 45.7 | 123.4 | 260.7 |
| T>ca | 2.5 | 1.5 | 4.2 | 2.6 | 1.5 | 1.3 | 1.8 | 1.2 | 0.3 | 0.6 | 1.3 | 1.5 | 1.4 | 1.9 | 1.8 | 3.1 |
| Mcs | 5.9 | 9.6 | 21.8 | 24.8 | 4.1 | 8.4 | 18.8 | 30.2 | 1.0 | 3.8 | 15.7 | 19.8 | 5.2 | 15.7 | 20.5 | 62.4 |
| mncs | 0.7 | 0.6 | 1.6 | 1.6 | 0.8 | 0.9 | 1.4 | 2.6 | 0.7 | 1.3 | 1.9 | 1.6 | 0.8 | 1.5 | 1.4 | 2.9 |
| Mean mjs mcs | 8.9 | 15.1 | 17.3 | 19.7 | 5.3 | 9.7 | 16.4 | 19.5 | 1.6 | 4.7 | 16.9 | 17.0 | 6.1 | 14.0 | 17.4 | 25.5 |
| Max mjs mcs | 23.2 | 61.4 | 105.6 | 187.9 | 13.8 | 34.1 | 75.7 | 111.5 | 3.6 | 18.9 | 98.1 | 121.8 | 18.4 | 52.9 | 111.7 | 180.3 |
| Mean mnjs | 1.0 | 0.9 | 1.2 | 1.3 | 0.8 | 1.0 | 1.2 | 1.6 | 0.8 | 1.2 | 1.5 | 1.2 | 0.9 | 1.2 | 1.2 | 1.3 |

Table 4 shows the mean indicator scores within the four clusters for each discipline. The majority of indicator scores increase across the clusters and it is the same indicators that increase in value in all fields. Analysis of the mean values confirmed that the indicators percent self-citations (*%sc)* and percent not cited *(%nnc)* decreased across clusters, that is the low scoring researchers in Cluster 1 have the highest proportion of un-cited papers and self-citations. This was expected as the composition publications and citations used to calculate these ratio indicators increases from Cluster 1 to Cluster 4. Across all disciplines PhD students and Post Doc researchers were dominant in Cluster 1, and seniority increased in Cluster 2 and 3 along with the academic age of the researcher. Consequently, *hnorm* generally decreased across clusters, as this indicator calculates the proportion of productive papers, *h,* to the total number of papers a researcher has produced. As the researchers in Cluster 1 have fewer publications the ratio productive papers to total papers is smaller and vice versa for researchers in Cluster 4. However, because *hnorm* normalizes *h* in this way, it enables comparison across fields, (Levitt & Thelwall, 2014), and shows that the Philosophers in Clusters 1,2,3,and 4 have on average a very similar number of citations per paper as their cluster counterparts in Astronomy, Environmental Science and Public Health. The extreme indicator values in Cluster 4 were scored by associate and full professors who, across all fields, ranked the highest scores in *C, sc, AW, h, ħ, Cless5, Csc, sum pp top n cits, hg, Q2, h2 and Nprod.* Meaning that these researchers, although they do not necessarily publish the most, are cited the most, are cited quickly and have the most impactful papers even when adjusted to the performance of their entire portfolio.

The researchers in Cluster 3, the high scorers, come second to Cluster 4 researchers when ranked according to the *h-type indicators, (h, g, h2, Q2, hg,* etc*)* but scored higher on field normalized indicators such as *mncs, sum pp top prop* and *mean mjs mcs, max mjs mcs.* Further, they achieved a higher rank placement than Cluster 4 researchers when the *SIG* indicator was used i.e. the most significant paper, showing that they have one very high scoring publication, higher than citations to papers produced by Cluster 4. Cluster 2, the middle performers, scored well on indicators that normalized or rewarded authorship, *FracCPP, mean pp collab* and *mean pp int collab*. This cluster ranked generally lower than Cluster 3 across the *h-type* indicators but using the *m index,* i.e. the median number of citations per paper in the *h*-index, increased their rank position above Cluster 3, but below Cluster 4, researchers. The researchers with low scores, Cluster 1, have the fewest publications and citations and rank below the researchers in Clusters 2, 3 and 4 across *h-type* indicators. Removing self-citations or calculating *CPP* further reduced their scores, however using the *mg* or *m-quotient,* which normalize the *h* and *g* indices for the academic age of the researcher, or indicators of the currency of papers, *PI, Cless5* raised the scores on a par with the researchers in Cluster 2 and in some cases higher than Cluster 3.

*Likelihood of researcher placement in cluster*
In researcher assessment, the academic age and seniority of the researcher are important considerations for contextualizing bibliometric statistics. Table 4 shows publication count increased with cluster chronology and Table 5 presents cluster composition. To explore if Cluster 1, the cluster with the lowest publication count, was dominated by junior researchers, and consequently if Clusters 2, 3, and 4 were progressively dominated by senior researchers, we computed the odds of a researcher who belongs to a particular seniority belonging to Cluster 1, 2, 3 or 4. SPSS crosstabs and risk statistics were used for this analysis. Table 6 presents the likelihood for seniority classes belonging to a cluster. The interpretation of Table 6 is this: An odds ratio of 1 means that the seniority has a similar likelihood to other seniorities to belonging to a cluster. The larger the odds ratio the more likely this event is expected to occur, odds ratio less than 1 are interpreted to indicate a protective effect, meaning that the seniority is less likely to be in this cluster. An odds ratio cannot be used to assess the causation of seniority and the resulting clustering, but it can help us determine if seniority influences the composition of clusters.

**Table 5**: Cluster composition. Field, academic age in years and share of researchers by academic seniority.

| Field | Age, years (range) | Seniority |
|---|---|---|
| *Astronomy* | | |
| Cluster 1 | 7 (2;18) | 12/15 Ph.D students, 29/48 post doc, 8/26 assis. prof, 4/66 assoc. prof |
| Cluster 2 | 21 (10;33) | 6/48 post doc, 7/26 assis. prof, 31/66 assoc. prof, 14/37 professors |
| Cluster 3 | 15 (3;34) | 3/15 Ph.D students, 13/48 post doc, 11/26 assis. prof 20/66 assoc. prof, 15/37 professors |
| Cluster 4 | 23 (11;33) | 11/66 assoc. prof, 8/37 professors |
| *Environmental Science* | | |
| Cluster 1 | 9 (2;31) | 3/3 Ph.D students, 14/17 post doc, 18/39 assis. prof, 21/85 assoc. prof, 7/51 professors |
| Cluster 2 | 17 (7;36) | 3/17 post doc, 19/39 assis. prof, 44/85 assoc. prof, 20/51 professors |
| Cluster 3 | 22 (9;34) | 2/39 assis. prof, 20/85 assoc. prof, 23/51 professors |
| Cluster 4 | 32 | 1 professor |
| *Philosophy* | | |
| Cluster 1 | 9 (0;33) | 7/8 Ph.D students, 16/22 post docs, 29/43 assis. prof, 48/74 assoc. prof, 32/75 professors |
| Cluster 2 | 15 (3;33) | 1/8 Ph.D students, 5/22 post doc, 12/43 assis. prof, 19/74 assoc. prof, 35/75 professors |
| Cluster 3 | 22 (15;30) | 1/22 post doc, 2/43 assis. prof, 6/74 assoc. prof, 5/75 professors |
| Cluster 4 | 18 (7;33) | 1/43 assis. prof, 3/75 professors |
| *Public Health* | | |
| Cluster 1 | 9 (0;25) | 8/9 Ph.D, 12/14 post doc, 18/30 assis. prof, 15/50 assoc. prof, 6/29 professor |
| Cluster 2 | 15 (4;34) | 1/9 Ph.D, 2/14 post doc, 9/30 assis. prof, 26/50 assoc. prof, 10/29 professor |
| Cluster 3 | 21 (10;33) | 3/30 assis. prof, 8/50 assoc. prof, 10/29 professor |
| Cluster 4 | 22 (19;28) | 1/50 assoc. prof, 3/29 professors |

Table 6. percent within seniority within the 4 clusters and odds ratio associated with each seniority

| Astronomy | Cluster 1 | | Cluster 2 | | Cluster 3 | | Cluster 4 | |
|---|---|---|---|---|---|---|---|---|
| | % within seniority | Odds | % within seniority | Odds | % within seniority | Odds | % within seniority | Odds |
| Ph.D | 80 | 13 | - | - | 20 | 0.5 | - | - |
| Post Doc | 60 | 7.6 | 12 | 0.2 | 27 | 0.7 | - | - |
| Assis | 30 | 1.1 | 27 | 0.8 | 42 | 1.6 | - | - |
| Assoc | 5.3 | .07 | 46 | 3.4 | 33 | 1.0 | 15.8 | 2.9 |
| prof | - | - | 37 | 1.4 | 37 | 1.2 | 25 | 4 |
| **Environmental Science** | | | | | | | | |
| Ph.D | 100 | - | - | - | - | - | - | - |
| Post Doc | 82 | 12 | 17 | 0.2 | - | - | - | - |
| Assis | 46 | 2 | 48 | 1.2 | 5 | 0.1 | - | - |
| Assoc | 24 | 0.5 | 51 | 1.7 | 23 | 1 | - | - |
| prof | 13 | 0.2 | 39 | 0.7 | 45 | 4.5 | 2 | - |
| **Philosophy** | | | | | | | | |
| Ph.D | 87.5 | 4.9 | 12.5 | 0.2 | - | - | - | - |
| Post Doc | 72.7 | 1.9 | 22.7 | 0.5 | 4.5 | 0.6 | - | - |
| Assis | 65.9 | 1.4 | 27.2 | 0.7 | 4.5 | 0.6 | 2.3 | 1.3 |
| Assoc | 65.8 | 1.4 | 26 | 0.6 | 8.2 | 1.5 | - | - |
| prof | 42.7 | 0.3 | 46.7 | 2.6 | 6.7 | 1.0 | 4.0 | 6.0 |
| **Public Health** | | | | | | | | |
| Ph.D | 88.9 | 11.2 | 11.1 | 0.2 | - | - | - | - |
| Post Doc | 85.7 | 9.0 | 14.3 | 0.2 | - | - | - | - |
| Assis | 60 | 2.2 | 30 | 0.7 | 10 | 0.5 | - | - |
| Assoc | 30 | 0.3 | 52 | 2.9 | 16 | 1.0 | 2.0 | 0.5 |
| Prof | 20 | 0.2 | 34.5 | 0.9 | 34.5 | 4.4 | 10.3 | 11.7 |

Stand out observations are: in Astronomy, PhD students are 13 times as likely to score low values and be grouped in Cluster 1 than other seniorities, Associate Professors are 3.4 times as likely to be in Cluster 2 while the odds of a Professor being grouped in Cluster 4, the extreme high values, are 4 times as likely than compared to the other seniorities. In Environmental Science post-doctoral students are 12 times as likely to be in Cluster 1 and Professors 4 times as likely to be in the high scoring Cluster 3. The pattern is similar in Philosophy and Public Health: PhD students are respectively 5 and 11 times as likely to be in Cluster 1 and Professors 6 and 11.7 times as likely to be in Cluster 4. In summary, junior researchers (PhD students and Post Docs) are more likely to be in Cluster 1, the likelihood of assistant and associate professors being placed in specific clusters unclear, and Professors are most likely to be in Cluster 3 or 4. The observed odds ratios could also be inaccurate because of selection bias, as our sample of researchers is by no means random.

*Ordinal logistic regression to predict cluster membership*

The obvious confounder associated with seniority and cluster placement, is the academic age of the researcher. In this paper academic age is the number of years since the researcher's first publication recorded in WoS, not the actual number of years they have been active as a researcher. Academic age is then a proxy for the coverage of the researcher in WoS. Coverage can distort the magnitude of the observed relationship between seniority and clustering which we already pointed out in the design stage of this study. To assess whether academic age is a confounder, we need to look at the association between cluster membership and seniority separately for academic age.

We ran an ordinal logistic regression to assess if academic age has a statistically significant effect on cluster membership. The regression analysis allows interpretation of the odds that academic age or seniority has a higher or lower value on cluster membership *if* the data meets the four assumptions that are required for ordinal regression to give a valid result: 1) cluster membership is a dependent variable, 2) the model contains five independent variable groups on an ordinal scale (the five seniorities) and the continuous variable academic age (in years, ranging from 1 to 36 years), 3) there is no strong multi-collinearity between academic age and seniority in Environmental Science and Philosophy PPMC 0.6, and Public Health PPMC 0.5. In Astronomy however multi-collinearity is present, academic age and seniority displayed a strong relationship PPMC 0.8. This leads to problems with understanding which variable contributes to the explanation of the cluster membership and technical issues in calculating an ordinal regression. 4) The assumption of proportional odds was met in Environmental Science, Philosophy and Public Health. The model was statistically significant and the test of parallel lines indicated the proportional odds assumption had not been violated and there was stability across the thresholds of the entire range of clusters from 1 to 4. In Astronomy the test of parallel lines indicated instability across cluster membership thresholds.

Astronomy failed to meet two of the four assumptions and is therefore excluded the regression analysis. Violation of these assumptions are quite common (Bornmann et al., 2011), and without further tests, such as a non-proportional odds model, we cannot draw any conclusions on the influence of academic age and seniority on cluster membership in this discipline.

The full model containing both seniority and academic age was statistically significant in Environmental Science, Philosophy and Public Health indicating that the model was able to distinguish the ordered nature of the clusters. Across all disciplines an increase in seniority did not statistically increase the odds of placement in a higher cluster. In Environmental Science the model explained between 21% (McFadden R square) and 42% (Nagelkerke R squared) of the variance in cluster placement. An increase in academic age was associated with an increase in the odds of placement in a higher cluster, with an odds ratio of 1.15 (95% CI, 1.091 to 1.194) Wald $X^2(1) = 32.75, p > 0.001$), meaning that for every additional year, researchers are 1.15 times more likely to increase in cluster placement, controlling for seniority. In Philosophy the model explained between 8% (McFadden R square) and 17% (Nagelkerke R squared) of the variance in cluster placement. Academic age made a unique statistically significant contribution to the model, though smaller prediction in cluster placement, ratio 1.1 (95% CI, 1.050 to 1.140) Wald $X^2(1) = 18.439, p > 0.001$). Finally in Public Health the model explained between 23% (McFadden R square) and 45% (Nagelkerke R squared) of the variance in cluster placement and academic age statistically increased the odds of a researcher being placed in a higher cluster, odds ratio 1.15 (95% CI, 1.195 to 1.094) Wald $X^2(1) = 24.85, p > 0.001$).

Because the proportional odds assumption was not rejected, the conclusion is that in this dataset, seniority is not statistically associated with cluster placement. But the model academic age-cluster placement explained just under 50% of the variance in Environmental Science and Public health, and in Philosophy the model fit was quite poor, explaining about 10% of the variation. Returning to Table 4, an increase in publications and citations in cluster placement is visible. Running a correlation analysis confirmed this positive relationship, Table 7, and a test of the difference of the correlation coefficients showed no real difference in the strength of the relationship between publications and cluster, and citations and cluster, $Z_{obs}$ falls within the boundaries of -1.96 and +1.96. Publication and citation count were transformed to logscales and their distribution approached normality

Table 7. The statistical significance of the difference between correlation coefficients, (Sig. = .000)

| Discipline | N | PMO *Plog10* | $Z_1$ | N | PMO *Clog10* | $Z_1$ | $Z_{obs}$ |
|---|---|---|---|---|---|---|---|
| Astronomy | 192 | .74 | .962 | 192 | .85 | .895 | 0.65 |
| Environmental Science | 195 | .78 | 1.04 | 194 | .89 | .915 | 1.27 |
| Philosophy | 222 | .66 | .802 | 191 | .87 | .906 | 1.05 |
| Public Health | 132 | .75 | .973 | 131 | .86 | .900 | 0.59 |

*Changes in within cluster rankings*

Even if researchers belong to the same cluster, they were ranked in a different order depending on the indicator. The following projections are illustrated using Astronomy researchers, because this disciplinary group provides a comparable amount of researchers in each cluster. Figure 2, top left, shows cluster analysis projection on the Fractionalized citations, *Fc*, and Fractionalized publications, *Fp*, axes. We observed that citation and publication count generally increased from Cluster 1 through 4. However, taking Cluster 4 as an example, we observed large differences in the ranked performance of these nineteen researchers. Researcher ID164 ranked in first place using *Fp* whereas ID162 ranked 82nd out of all of the 192 Astronomy researchers. ID183 also, Cluster 4, ranked first place using *Fc* while the other 18 researchers in Cluster 4 were spread between 2$^{nd}$ to 65$^{th}$ ranked position. However, when all the Astronomy researchers were ranked after *C*, the raw citation count, these 19 researchers in Cluster 4 were ranked in the top 19 positions. Figure 2, top right, shows the projection on the *C* and *Fc* axes. Whereas *C* favours researchers with a high total citation count, normalizing citations to the number of contributing authors, *Fc*, reduces the dominance of Cluster 4 researchers in the top rank positions, lifting researchers from the other clusters up the rankings. Generally researchers who collaborate with ≤3 partners move up the rank, while collaboration with ≥4 authors fall in rank position. Of course the amount of papers and citations the collaboration is distributed across further influences this trend.

Figure 2, middle left, shows the projection on the citation count, *C*, and h-index, *h*, axes. The distribution is linear and there are clear thresholds between the clusters. Within the clusters, the top researchers in Cluster 4 and bottom researchers in Cluster 1 are ranked in the same positions on both indicators, the rank changing by roughly ± 1 rank position. We found the ratio *P* to *h* (total publications divided by *h* value) determined rank position. We observed that if a researcher has a ratio *P:h* of ≥3 they fall in rank position, and if the ratio is <3 they gain in rank position. A flaw of the h-index may be its inherent size dependence, in which larger numbers of publications generally command higher h-indexes as discussed in (Costas et al 2010; Vinkler 2007; Van Raan 2006)

Figure 2, middle right, shows the projection of the *m* and *A*-index. These two indicators are interesting to compare as *m* takes the median of citations to articles in the *h* core and *A* takes the arithmetic mean. Results show *m* rewards researchers with higher rank placements than *A*. The researchers in Cluster 3 and 4 are roughly placed 6 positions higher on *m* compared to *A*, however Clusters 1 and 2 are raised up to 20 positions on *m*. A dependence ratio similar to *P:h* was found on the projection of the *m* and *A* axes.

Figure 2, bottom left illustrates indicators that promote researchers in Cluster 1, 2, and 3. The projection, bottom left, shows the *CPP* and *hnorm* axes. Researchers in Cluster 1 scored the lowest citations per paper, *CPP*, see Table 4, however it is these researchers that score higher values on *hnorm*, which like *CPP* is a method of computing citations per paper. Figure 3, bottom right, shows projections of the *h* and *m-quotient* axes. The *m-quotient* normalizes *h* for the academic age of the researcher and increased researchers in the lower clusters rank placement. For example ranking using the *m-quotient* promoted researcher ID4 (Cluster 3) from *h* rank position 104 to *m-quotient* rank position 1. This researcher has 25 publications, 529 citations, and *h* 15 and an academic age of 3 years. Similarly, ID 23 (Cluster 1) also has *h*15 and rose from *h* rank position 100 to *m-quotient* rank position 35. This researcher has a career of 8 years, 53 papers and 634 citations. Conversely, researcher ID114, *h*15 Cluster 2, fell in rank position from *h* rank 103 to rank position 153 using the *m-quotient*. ID114 has an academic age of 23 years, 50 papers and 575 citations. In this indicator, a large number of years appears to cause a reduction in rank position.

The above analyses were repeated for Environmental Science, Philosophy and Public Health, apart from Cluster 4 which in all disciplines was very small, ≤ 4 members. The distribution of researchers within Clusters 1, 2, and 3 and patterns of change in the rank position were the same as observed in Astronomy, also observed was the ratio *P:h* which determined a rise or fall in rank position in all disciplines.

**Figure 2**. Clustering in Astronomy. Projection on bibliometric indices axes

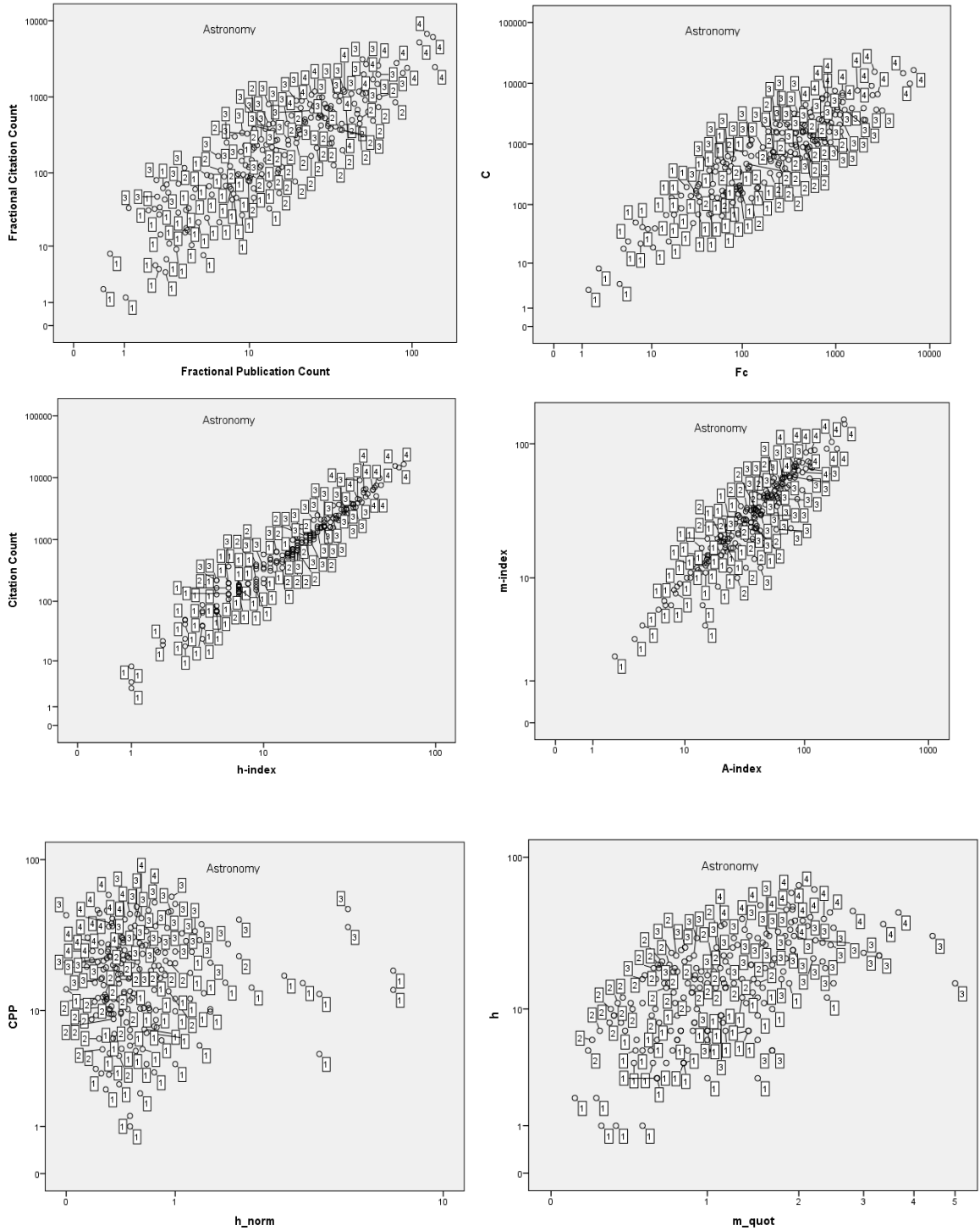

# Discussion and conclusion

This paper explores cluster analysis as a method to identify appropriate bibliometric indicators for groups of researchers. The cluster analysis was combined with an ordinal regression and odds analysis to investigate how 44 author-level bibliometric indicators present the academic performance of researchers, the likelihood of a researcher being placed in a specific cluster and appropriate indicators for different academic seniorities and disciplines. This investigation focused on European researchers in Astronomy, Environmental Science, Philosophy and Public Health.

*Statistical description of indicators*

The most important observation is that the amount of publications covered in WoS is fractional compared to the publications listed on researchers' CV. The indicators are thus a fractional representation of the effect of the researchers work or, if used in rankings, a limited depiction of the prestige of the researcher. Even though the underrepresentation of the researchers in WoS did not affect the computation of the indicators the resulting indicator values are considered misrepresentative of the researcher when compared to their profile on their CVs. The classic channel for bibliometric analysis is journal articles. Through this channel research is presented, discussed, critiqued and as a result validated. The participation of the individual researcher in this communication process is measured quantitatively by counting the number of a researcher´s published articles and the number of citations each article has received. The assumption is that to be included in journals the articles have passed peer-review and have received a stamp of "quality" and that articles are a common, constant type of scientific output, produced by researchers across all academic seniorities and articles that are cited more are more influential. Accordingly, counting articles and citations are equated to the production and visibility (or effect) of a researcher and have become standard bibliometric indicators used in the assessment of research performance. Already here, researchers who disseminate knowledge through other channels are at a bibliometric disadvantage – simply because output that is not counted becomes invisible. A major factor in the low bibliometric scores observed for Philosophy in this paper, is the lack of coverage of their work in WoS. The Philosophers in the study, listed in total 14,762 publications (articles, reviews, conference papers) on their CVs and we were able to identify 3,753 of these publications in WoS, approximately 24% of their papers. Likewise we identified 80% of the papers listed by the researchers on their CVs in Public Health, 58% in Astronomy and 47% in Environmental Science. Therefore researchers who are assigned a cluster membership based on low, middle, high or extreme scores can be biased by the lack of coverage of this discipline in WoS and hence the representation of the researcher through bibliometric indicators is also biased. As a result the perception of researcher performance could be distorted, clearly illustrating the importance of knowledge of the cultural, characteristic and contextual dimensions of the individual in both the application and interpretation of indicators (Bach, 2011; IEEE, 2013) as well as knowledge of the content of citation databases.

However, counting articles and citations is a verifiable and repeatable objective method so it continues to be used as base measures of research performance at the individual level that are combined using mathematical methods to produce a single integer number indicators that measure the productivity, currency, prestige and impact of the researcher. Accepting that author-level indicators are increasingly becoming institutionalized in national and university assessments, it is important the bibliometric community explore the disciplinary and seniority appropriateness of indicators, and come with educated recommendations for their application.

*Two step cluster analysis*

In this paper the two-step cluster analysis was explored as a method to identify disciplinary and seniority appropriate indicators. Like other clustering methods, hierarchical clustering can be criticized for producing arbitrary cluster solutions, as it has difficulty representing distinct clusters with similar expression patterns and as clusters grow in size, the expression patterns become less relevant. However, as argued in the Related Literature section, the two-step cluster approach was the logical choice of technique, that made sense and was rationally useful for the task (Schneider & Borlund, 2007a). The accuracy of the clusters is dependent on the thoroughness of the preliminary data preparation processes, also discussed in (Chawla, 2006; Su et al., 2013) hence in this paper preparation was extremely thorough. The clustering algorithm was successful in identifying 4 groups of researchers that were substantively interpreted as low, middle, high and excessively high performers. Our confidence in the method must not override validating the proposed cluster solution with common sense observations of the variables that make up the dataset and what the data represents, as test statistics do not always lead to definitive decision on the number of clusters and more than one solution can be appropriate. Goodness of fit and likelihood statistics are a *guide* to selecting the number of clusters and should not be adhered to automatically. The common sense approach used in this paper is not to rely entirely on statistical clustering parameters, but to verify cluster composition and indicator scores to researcher CVs and select the most appropriate indicators according to these comparisons. Interpretation of within and between cluster similarities and dissimilarities based on statistics alone was uninformative and was supported using other methods, likewise interpretation of the bibliometric scores and hence characteristics of researcher performance was only insightful when supported by information from the researchers CVs. If the analysis resulted in the correct division based on indicators is uncertain, as we do not know the ground truth, though it does provide a sensible solution.

*Validity of the clusters*
Logically, researchers belonging to different cluster groups are, according to the grouping based on indicator scores, very different from each other. A substantive interpretation of the clusters was very simple and fulfills the interoperability criteria for a good classification, (Bacher, 1996). Likewise the clusters displayed stability, removing a small number of cases (though not the outliers) did not change the resulting clusters, and the clusters displayed relative validity as the classification was better than the null model which assumes no clusters are present in the data. Previous studies have investigated and discussed the differences between junior and senior researcher performance on indicators (Costas et al 2010) and likewise differences between and within academic domains (Claro and Costa 2011; Archambault and Gagné 2004) so the resulting hierarchical division between clusters was to be expected. However, usefulness of further validation tests in respect to the quality of the clusters identified in this study is questioned, as the clustering was based on a hierarchical method of partitioning. Common measures such as the Rand Index are only valid for strict partitionings. The cut-off point between clusters of researchers based on the ground truth would have to be determined, which we do not have. The Two-step cluster belongs to a family of exploratory data analysis techniques called unsupervised learning, where there is no error or reward signal to evaluate potential solutions as there is no way to evaluate the stability of the algorithm. There could be a bias in the SPSS clustering algorithm towards partitions that are in accordance with a certain clustering criterion and a different algorithm in another statistical program like SAS or R could have provided a different solution(s). The heterogeneity between clusters was tested by comparing cluster averages per indicator, and in the majority of cases there was a statistical difference between clusters, but this is not interpreted as necessarily an important difference between groups of researchers. The advantage of using mean cluster scores is they are less susceptible to noise and outliers.

*Composition of the clusters*
The strength of the method used in this paper is the further exploration of the proposed automatic clustering of researchers, the investigation of the role of indicators in cluster composition and the likelihood of researchers being placed in a cluster and the rich dataset permits comparison of researcher scores with publication activity listed on researchers' CVs. Within each discipline the researchers were grouped into 4 different clusters: Cluster 1 low indicator values, Cluster 2 middle, Cluster 3 high and Cluster 4 extremely high indicator values. In all fields, researchers placed in Cluster 1 proportionally cited themselves the most, had the highest proportion of un-cited publications but benefitted in an increase in rank placements by indicators that measured the currency of their work and the ratio of productive papers they have in their portfolio. The results were due to the ratio of small amount of publications to the youth of these publications rather than specific publication and citation characteristics of the research field, and further the dominate quantity of junior researchers in this cluster.

Dependent on discipline, a Cluster 1 researcher had the average academic age between 7 and 9 years, and was typically a PhD student or Post Doc researcher. In Cluster 2 researchers had an academic age between 15 and 21 years and were assistant or associate professors. Similarly Cluster 3 researchers had an academic age of 15-21 years and were identified as associate or full professors. Researchers in Cluster 4 were not vastly older, having an academic age of 18-23 years and were associate or full professors. The production and citation profile between clusters was though vastly different where researchers in Clusters 1, 2, 3, and 4 had between, respectively, 6-17 (5.6-150), 24 -53 (65.5 -741), 35-140 (432.7-2602.9) and 99-425 (1971.2-14.141) publications and (citations). The difference in researcher profiles and citations between clusters in Astronomy was not as clean cut as in the other disciplines and it was not possible to conduct a valid regression. In Astronomy the researchers documented on their CVs that they worked in vast author-collaborations. This collaboration means that junior researchers can quickly build a relatively substantial portfolio and have highly cited publications, and thus ranking researchers in groups determined by age and seniority do not make sense, as in the collaborative world of Astronomers this distinction is fluid.

*The importance of each indicator as a predictor of the cluster*
Previous studies have shown that it does not make sense to compare researcher rankings and indicator scores across disciplines (REF; REF) hence the interest in producing indicators that are specifically designed for cross-disciplinary comparisons, such as the *hnorm* (Sidiropoulos et al., 2007), *x*-index (Claro & Costa, 2011) and $\pi$-index (Vinkler, 1996). The results in this paper show that different indicators were stronger in ranking authors and determining cluster placement in different disciplines: in Astronomy the *h2* indicator, Environmental Science *sum pp top prop,* Philosophy *Q2* and Public Health *e*. Four very different indicators for four very different disciplines, that are designed to capture different aspects of researcher performance, respectively cumulative achievement (*h2*), papers at the top of the field (*sum pp top prop)*, effect of all productive papers (*Q2*), and, production and effect of highly cited papers (*e*), Appendix 1. These indicators are calculated using different mathematical equations that can be rationalized to better fit the publication and citation characteristics of each discipline, e.g. *h2* corrects for the ratio many papers to few extremely highly cited papers that is a common characteristic of output in Astronomy to produce a more granular and comparable indication of a researcher's productivity and impact. On the other hand it could be argued that the role of these indicators in defining the clusters does not in fact define the publishing characteristics of researchers and hence identify disciplinary appropriate indicators. Instead, these indicators mathematically combine bibliometric data that produce a better fit to the *F* statistic that was used to predict the statistical importance of the indicator for cluster membership. This consideration was explored by ranking the researchers by the predicted indicators of importance and it appeared that the sole importance of these indicators was to produce clear thresholds between groups of researchers in the cluster algorithm and thus demarcate the four clusters, i.e. these indicators grouped the researchers into four definite clusters, *F* =1, whereas ranking the researchers with indicators with weak prediction strength, *F* <1, produced muddled groupings. Thus it follows that interpretations of researcher performance as low, middle, high or extremely high based on rankings is not informative, as one would be interpreting the computer model and not actual disciplinary and individual profiles.

But why is the clustering algorithm using different indicators in different disciplines to group the researchers, and can we use knowledge of the researchers' curriculum vitae to explore if a cluster analysis can say *something* about the academic performance of researchers rather than the mathematical performance of the clustering algorithm? Although the clustering algorithm differentiates between researchers using indicators that are purely of statistical importance rather that scientific importance, it was possible to make some interesting observations about cluster placement based on the dominate indicators used to form the clusters. Researchers in Cluster 2 displayed *mean pp collab* and *mean p pint collab* at the same or only slightly lower level than Clusters 3 and 4. The *%sc* and *%nnc* was higher than Cluster 3 and 4 but lower than Cluster 1 and combined with their low scores on *mncs, T>ca* and high scores on *NprodP* we deduced that these researchers generally produce papers that perform better than the expected performance of articles in the sources they choose to publish in, but on a field level are cited less than average, *mncs*. It was very difficult to distinguish characteristics of Cluster 2 researchers using the bibliometric indicators, and studies before ours have commented, the indicators were the most useful in identifying differences between top and bottom researchers rather than the middle set where the relative ranking of researchers is significantly altered, i.a. Meho and Yang (2007). Continuing the analysis of the *mncs* indicator, we observed that Cluster 1 researchers were cited between 40 and 10% less than the expected field average and Cluster 3 and 4 researchers cited between 40 and 90% more than expected. Especially Philosophers performed well on this WoS indicator. The noticeable difference between Clusters 2 and 3 was the jump in the average citation score of the journals in which the researchers publish, that is a marked difference in the indicators of prestige, such as the *mcs, mncs, max mjs mcs,* the h-type indicators and *SIG*. Generally the high scores in Cluster 3 and the outliers in Cluster 4 were the researchers that scored well on indicators of impact, like the h-type indicators but they did not necessarily collaborate more (*APP, mean pp collab*) or display greater cognitivity (*mean p pint collab*) with their peers. The smallest cluster, Cluster 4 the outliers, excelled in the number of publications, were cited the most over their entire body of work, and wrote articles that were more productive than disciplinary averages. This does not necessarily mean that they are cited more than expected, as the *T>ca* indicator shows, these scores were field dependent and varied between clusters. There are also too few researchers in Cluster 4 to draw any conclusions based on statistical analysis or draw conclusions on the observed trends in the dataset, Table 5.

The analyses of the Philosophy researchers was hampered by the lack of coverage in WoS, that as a result provides distorted indicator values that do not provide useful information. Focusing only on the percentage based indicators to enable comparison within and between clusters, inter-institutional collaboration (*mean pp collab*) was lower in Clusters 1 and 2 than in 3 and 4. Not surprisingly these researchers scored low on cognitivity, *mean p pint collab,* as this indicator shows the proportion references in the publications linking to other WoS publications. Philosophers scored the lowest *%sc* compared to the other fields, with the amount increasing from clusters 1 to 4 whereas in the other fields the average *%sc* value decreased. Conversely *%nnc* decreased from cluster 1 to 4 and displayed the same trend as in the other fields: researchers in Clusters 3 and 4 had the fewest non-cited documents. Philosophy researchers also had the lowest percentage of publications cited within 5 years of their publication however the overall trend was the same as in the other fields, in that researchers in Clusters 1 and 2 had the proportion most recent citations to all publications. However, scores on the effect and prestige indicators indicated that citations were not as readily given in Philosophy as in the other fields, which we interpret as the value of one citation in Philosophy is worth more than the other fields, a useful consideration if we had to compare disciplines across citation indicators. Not all indicators displayed a statistical difference between clusters, Table 3, therefore the clusters are heterogeneous on only a small proportion of the 44 indicators, between 16 and 31 dependent on the discipline. Noticeably field normalized indicators do not contribute statistically to the clusters, which contradicts the observations of researcher performance in Cluster 3. Interestingly the indicators that did not display a statistical difference between-clusters thresholds were publication-based indicators: indicators of production (*P, Fp*), indicators that count collaboration without adjusting for age (*APP, mean pp collab, mean p pint collab, POPh*), and indicators that normalize the number of citations across all publications (*CPP, Fc, FracCPP*). Common across all fields, the *h-type* indicators contributed most strongly to predicted cluster solutions.

*Likelihood of researcher placement in cluster*
The Clusters contained a muddle of different academic seniorities, yet the odds analysis clearly demonstrated that PhD students and Post Docs had the greatest odds of being placed in Cluster1 and professors in Cluster 4, but the odds placement of assistant professors and associate professors was uninformative, as previously discussed in the section *composition of clusters*. The ordinal regression confirmed that seniority did not contribute to cluster formation, rather academic age was statistically significant for cluster placement with a substantial increase in the odds of researchers with a longer publication history indexed in WoS (academic age) being placed in higher clusters (15% increase with each unit increase in age). But this did not explain all of the variance in cluster placement, and the $Z_{obs}$ analysis confirmed the strong influence of publications and citations, which together with academic age is suspected to steer the placement of researchers in the clusters.

**Limitations**
The role of the statistically significant indicators is interpreted cautiously, as the data integrity could be compromised due to missing data, outliers, chance and that the predictors of importance have been chosen by the computer, not by us, the researchers. Even though the scores on indicators between clusters differ significantly, we make no assumption that the difference important, as statistical significance is not necessarily the same as scientific significance. The results are based on *p* values, thus are unlikely to replicate in another sample. Limitations: Common issues affecting all clustering analyses of bibliometric indicators, is the completeness of the data sources used to compute the indicator. As performance indices behave differently across data sources and the interpretation of optimal number of clusters arbitrary without strong methodological arguments and quality control of the data (Liu, 2009).

# Conclusion
In this exploratory study, we used a two-step cluster methodology to study 44 bibliometric author-level indicators with the aim to recommend disciplinary and seniority appropriate metrics. The results show that different indicators were stronger in different disciplines in ranking authors as low, middle, high and extremely high performers: in Astronomy the *h2* indicator, Environmental Science *sum pp top prop,* Philosophy *Q2* and Public Health *e*. Four very different indicators for four very different disciplines, that are designed to capture different aspects of researcher performance, respectively cumulative achievement (*h2*), papers at the top of the field (*sum pp top prop)*, effect of all productive papers (*Q2*), and, production and effect of highly cited papers (*e*). It was not possible to give definite recommendations of seniority appropriate indicators. Seniority did not statistically contribute to cluster placement, the clusters were generally a muddle of different seniorities. It is suspected that cluster placement was determined by a ratio between the researcher's academic age, number of publications and number of citations - three variables directly influenced by the researcher's coverage in WoS. Thus the coverage of the individual researcher and indexing policies of the citation index used to collect the bibliometric data highly influenced indicator scores. Accordingly a disconnection between the prestige of the researcher reported on their curriculum vitae and the prestige indicated by the calculated indicators was observed and consequently the perception of the researcher based on indicator values, can result in a misconception of actual researcher performance. For example the *hnorm*, *mg* and *m*-indices promoted Cluster 1 researchers up rank placements ahead of Cluster 3 and 4 researchers. Within cluster hierarchy showed the rank position of the researcher was predominantly determined by the ratio *P:h*-index. *P:h* of ≥3 was linked to a fall in rank position, and <3 a gain in rank position, which means that researchers can strategically include or exclude publications from the calculation of the indicators to improve rank position and artificially plump their statistics.

The two-step cluster approach was explored as an applicable method to identify disciplinary and seniority appropriate indicators. It is not possible to give a definite answer if the approach is most appropriate in all circumstances. All ordinations are wrong to some extent and introduce bias into the solution. The clustering identified different bibliometric indicators that were more appropriate in some fields than others and was successful in creating a sensible foundation for the further analysis of indicators supplemented by other statistical methods and contextual information from researcher CVs. Even though the results of our study cannot be generalized outside of our dataset, it is important to do studies like ours that critically investigate the usefulness of statistical models and the application of bibliometric indicators. This will help us learn more about the advantages and disadvantages of quantitative analysis of researcher performance and perhaps help us illuminate the inappropriate application of methods and stop creating superfluous indicators.

## Acknowledgements

This work was partially funded by ACUMEN (Academic Careers Understood through Measurement and Norms), FP7 European Commission 7th Framework "Capacities, Science in Society", grant Agreement: 266632. Opinions and suggestions contained in this article are solely the authors and do not necessarily reflect those of the ACUMEN collaboration.

Appendix 1: 44 bibliometric indicators of individual performance. The columns, from left to right, present the full name of the indicator, the abbreviation, the definition and the aim of the indicator, as proposed by the creator of the indicator.

| ID | Indicator | Abbr. | Definition | Aimed to assess |
|---|---|---|---|---|
| 1 | Number of publications | P | Total number of publications by the researcher | Production |
| 2 | Fractional Publications | Fp | Each publication divided by number of authors, limited to max. 10 authors | Production if the author had worked alone |
| 3 | Authors-per-paper | App | average number of authors per paper over all papers | Collaboration |
| 4 | Mean pp collaboration | Mean pp collab | Percentage inter-institutional collaboration type, taken from author byline information in Web of Science. | Collaboration |
| 5 | Mean pp internal collaboration | Mean pp int collab | The proportion of cited references in the publication linking to other WoS publications. A paper with an internal coverage of 0.8%, means that 80% of the references of this paper are covered by the WoS (since 1980) | Cognitivity |
| 6 | Number of Citations | C | Total number of citations received by publications of the researcher (including self-citations) | Effect of production |
| 7 | Citations minus self-citations | Csc | Total citation count, self-citations removed | Citations from external parties |
| 8 | Number of self-citations | sc | Sum of self-citations | Building on own research |
| 9 | Percent self-citations | %sc | Number of self-citations divided by total citations | Identifies unwarranted self-promotion |
| 10 | Number not cited | nnc | The sum of uncited papers | Non-effectual papers |
| 11 | Percent not cited | %nc | Share of uncited publications | Percentage of work that has not been cited to the present date |
| 12 | Citations per paper | CPP | The average number of citations per paper, C/P. | Average effect per paper |
| 13 | Most Significant paper | SIG | The paper with the highest number of citations | Most effectual paper in researcher's portfolio |
| 14 | Citations less than 5 years old | Cless5 | Number of citations less than 5 years old, from the publication of the paper. Publication year is Zero | Currency of citations |
| 15 | Price Index (Price, 1970) | PI | Percentage references to documents, not older than 5 years, at the time of publication of the citing sources | Currency of citations |
| 16 | Citation age (Egghe & Rousseau, 2000) | Cage | Mean difference between the date of publication of a researcher's work and the age of citations referring to it. | Currency of citations |
| 17 | Fractional citation count | Fc | Gives an author of an m-authored paper only credit of c/m if the paper received c citations | The effect of each author of a paper |
| 18 | Fractional Citations per Paper | FracCPP | Fc/Fp | The average effect of each per paper, adjusted for the numbers of author per paper |
| 19 | Sum pp top number of citations | Sum pp top n cits | Proportion papers that receive more than 10 citations. 1 is that the paper has more than 10 citations and 0 that is has less | Productivity and impact of a researcher |
| 20 | h-index, (Hirsch, 2005) | h | Publications are ranked in descending order after number of citations. Where number of citations and rank is the same, this is the h index | Productivity and impact of a researcher |
| 21 | Age Weighted Citation Rate (Harzing, 2012) | AWCR | the number of citations to a given paper divided by the age of that paper. Sum over all papers | Productivity and impact allowing younger, less cited papers to contribute to the index |
| 22 | Per-author AWCR, (Harzing, 2012) | AWCRpa | AWCR normalized for the number of authors for each paper | The per-author age-weighted citation rate is similar to the plain AWCR, but is normalized to the number of authors for each paper. |
| 23 | g-index (g), (Egghe, 2006) | g | Publications are ranked in descending order after number of citations. The cumulative sum of citations is calculated, and where the square root of the cumulative sum is equal to the rank this is g-index | Productivity and impact of a researcher, including highly cited papers |
| 24 | h2, (Kosmulski, 2006) | h2 | Weights most productive papers by finding the cube root of all citations (not just citations to h-core articles). | Productivity and impact of a researcher, including highly cited papers |
| 25 | POP h (Harzing, 2008) | POPh | Divides number of citations by number of authors for that paper, then calculates the h-index of the normalised citation counts | Productivity and impact of a researcher, if the researcher had worked alone. |
| 26 | m-quotient, (Hirsch, 2005) | m-quot | h divided by academic age | Productivity and impact of a researcher normalized for academic age of researcher |
| 27 | mg-quotient | mg-quot | g divided by academic age (Egghe 2006) | Productivity and impact of a researcher normalized for academic age of researcher |
| 28 | Q2 (Caberizoa et al., 2012) | Q2 | Q2 is the geometric mean of h-index and the median number of citations received by papers in the h-core | Productivity and impact of a researcher. Relates the number of papers to the impact of these papers in the h-core |
| 29 | Age Weighted h, (Harzing, 2012) | AW | Square root of AWCR, suggested as comparable to the h index | Productivity and impact of researcher, normalized for academic age of researcher |
| 30 | e-index, (Zhang, 2009) | e | The e-index is the (square root) of the surplus of citations in the h-set beyond $h^2$, i.e., beyond the theoretical minimum required to obtain a h-index of 'h'. | Supplement to the h index. Production and effect of highly cited papers, |
| 31 | M index (Bornmann et al., 2008) | m | Median number of citations received by papers in the h-core | Supplement to the h index. Median number of citations to core papers |
| 32 | A index (Jin, 2006; Rousseau, 2006) | A | Average number of citations in h-core thus requires first the determination of h. | Supplement to the h index. Mean number of citations to core papers |
| 33 | AR index (Jin et al., 2007) | AR | the square root of the sum of the average number of citations per year of articles included in the h-core, as such the AR index can decrease over time. | Supplement to h index. Accounts for the actual number of citations and age of most productive papers. |
| 34 | Hg (Alonso et al., 2010) | hg | Square-root of (h multiplied by g) | Compare researchers with similar h and g indexes. |
| 35 | Normalized h, (Sidiropoulos et al., 2007) | h-norm | Normalized h=h/np, if h of its np articles have received at least h citations each, and the rest (np-h) articles receive no more than h citations. | Normalizes h-index to compare scientists across fields. |
| 36 | ℏ, (Millers h) (Miller, 2006) | ℏ, Millers_h | Square root of half the total number of citations to all publications | Comparison across field and seniority of papers in the productive core |
| 37 | Sum pp top prop | Sum pp top prop | Proportion of papers in the top 10% of the world. 100% means that the article belongs to this set of papers, 0 means not. | Identify researcher's papers that are rated top of their field |
| 38 | Number of productive papers (Antonakis and Lalive 2008) | NprodP | Based on the calculation of the Index of Quality and Productivity, which is a bench mark computed using the number of years since the researcher defended his/her doctorate, the number of papers published, times cited and the top three areas in which the researcher is cited. NprodP is the number of papers that perform better than the benchmark | Amount of papers that are cited more frequently than average papers in the researcher's specialty |
| 39 | Times cited more than average (Antonakis & Lalive, 2008) | T>ca | Based on the calculation of the Index of Quality and Productivity and the NprodP indicators: T>ca is the rate the NprodP papers (adjusted papers) perform better than average | How much more than average, as a ratio, the researcher is cited |
| 40 | Mean citation score | mcs | mean citation score (journal) self cites not included | Journal impact (prestige of journal the researcher publishes in) |
| 41 | Mean normalized citation score | mncs | Relates article to world average in regards to document type, publication year and field. 0.9 means cited 10% below average, 1.2% cited 20% above. | Mean normalized citation score (adjusts for field, article type and publication year. SC not included) |
| 42 | Mean journal score : mean citation score | Mean mjs mcs | mean citation score of all publishing journals the researcher has published in. | Prestige, benchmark. Expected number of citations of the articles in journals the researchers publish in. |
| 43 | Maximum journal score mean : citation score | Max mjs mcs | Highest citation score of a journal the researcher has published in | Prestige, most significant place of publication |
| 44 | Mean normalized journal score | Mean mnjs | Average impact of the journals in which the researcher has published compared to the world citation average in the same subfields | Prestige, corrects for differences among fields |

Appendix 2. Boxplots of each bibliometric index, log10-transformed

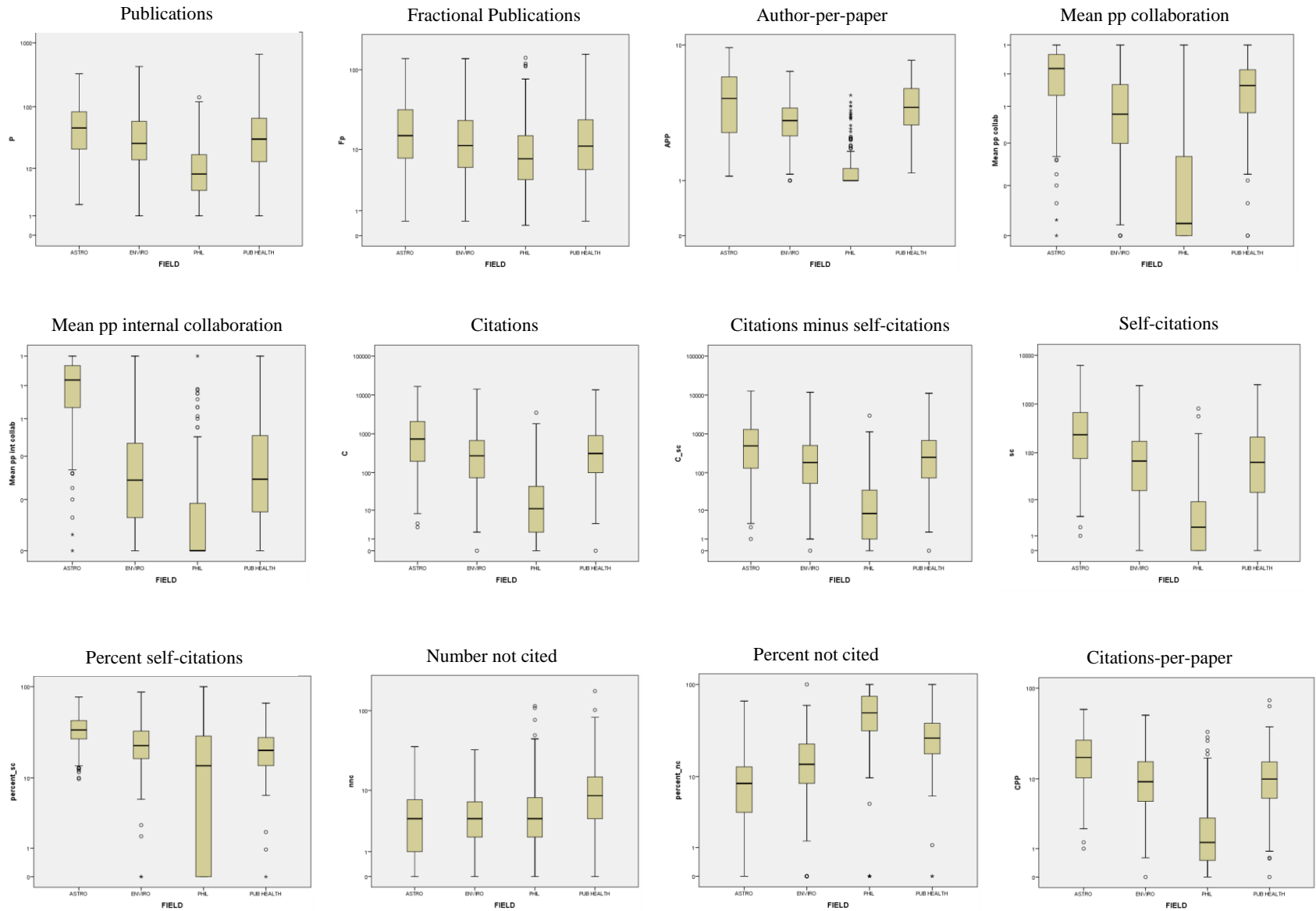

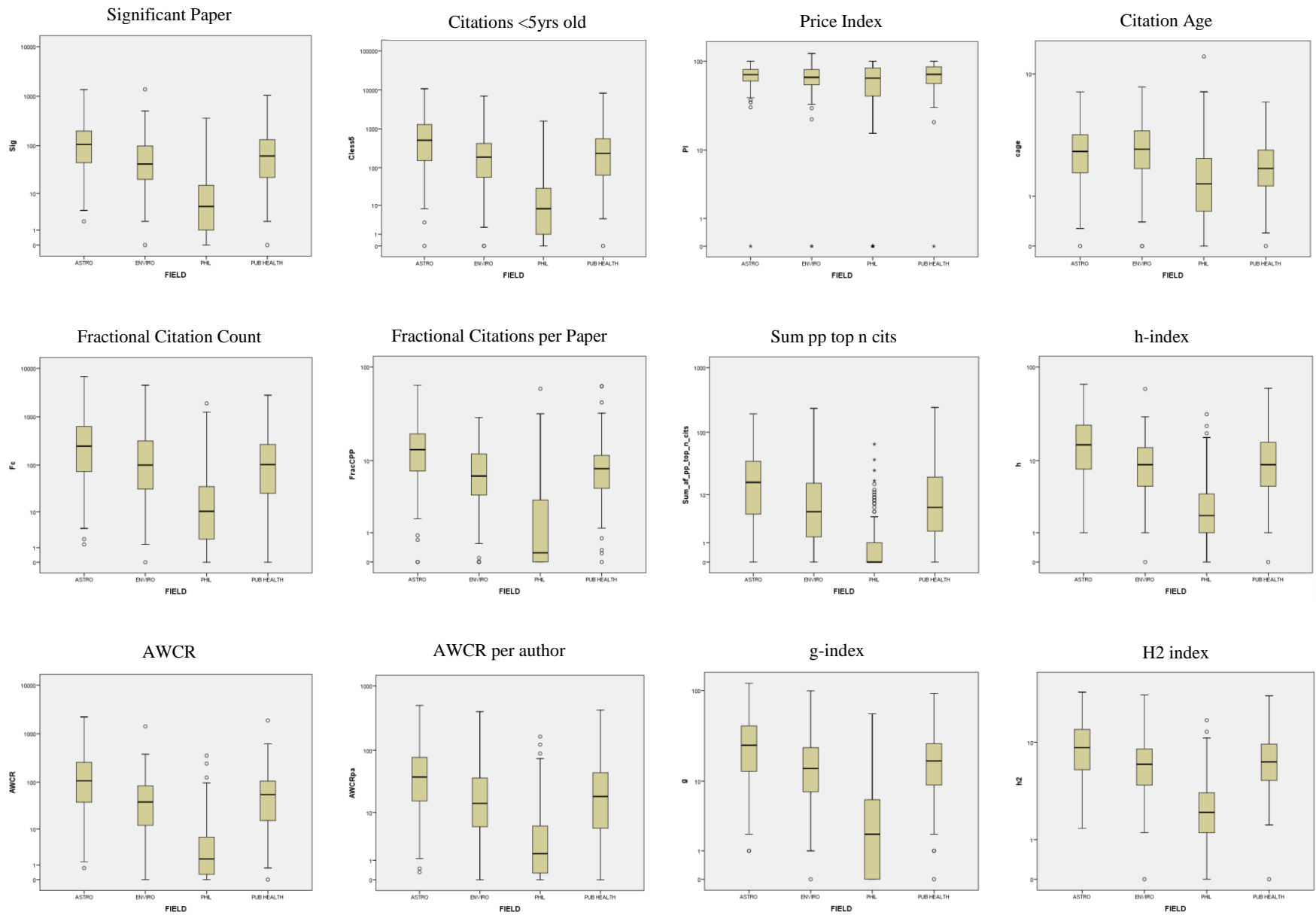

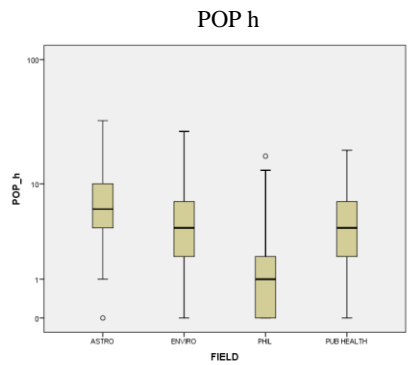
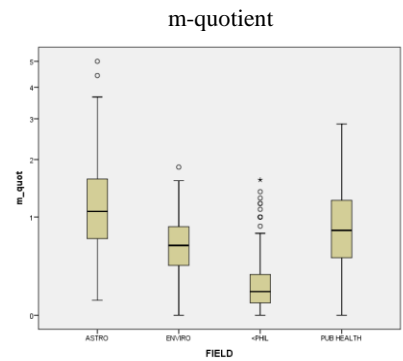
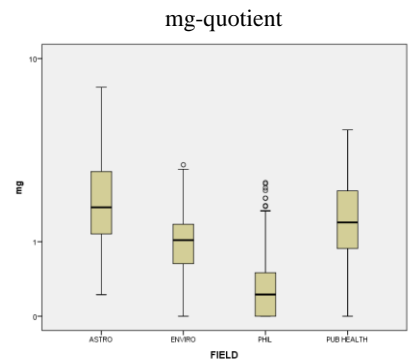
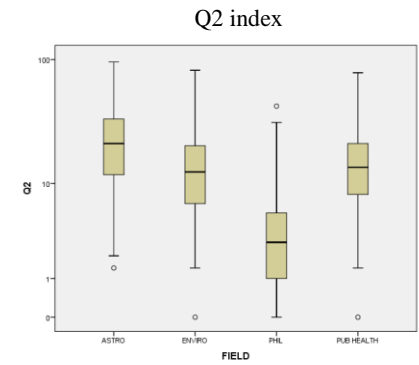
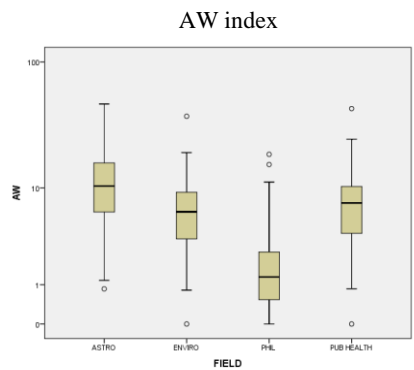
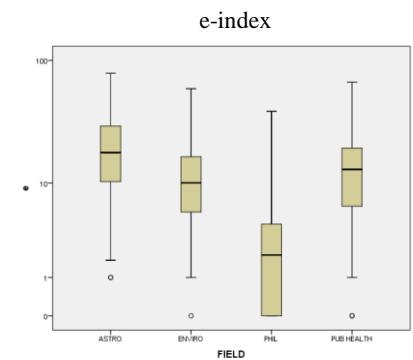
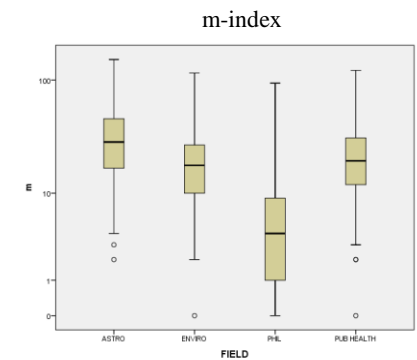
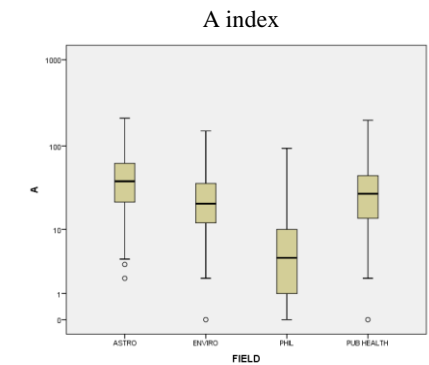
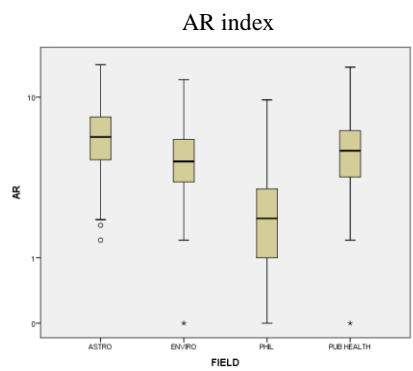
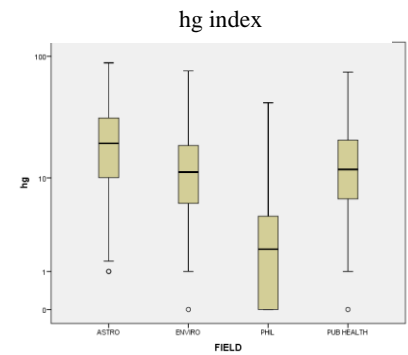
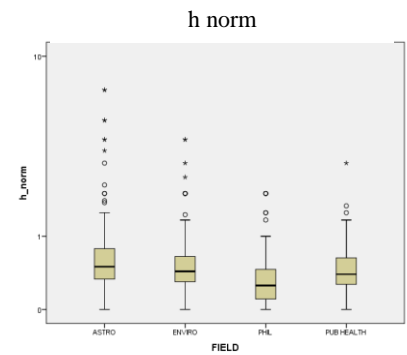
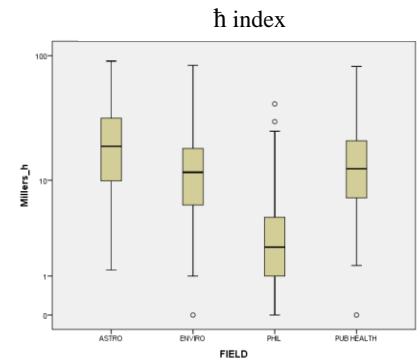

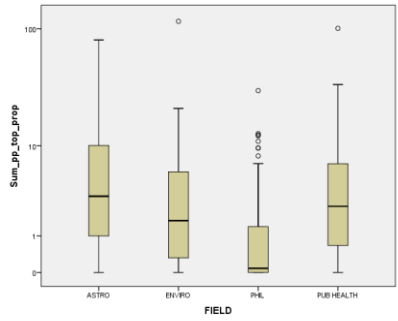 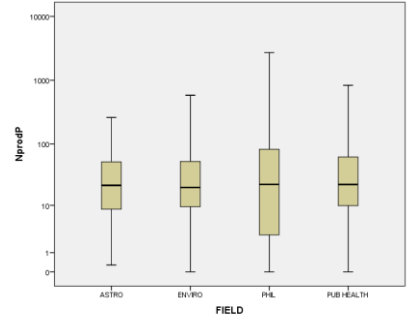 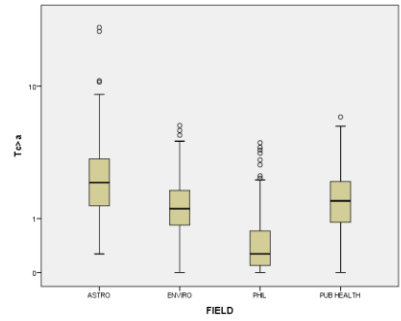 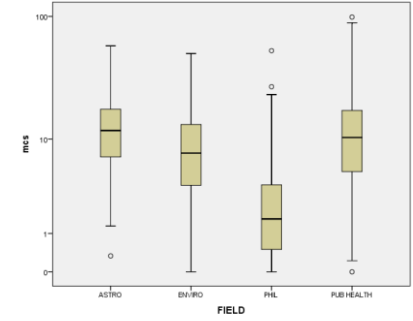
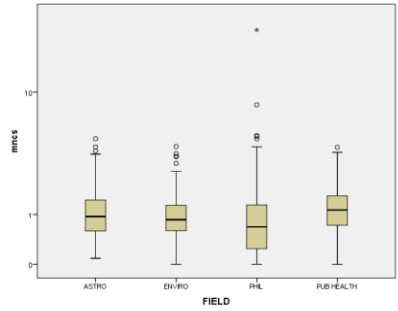 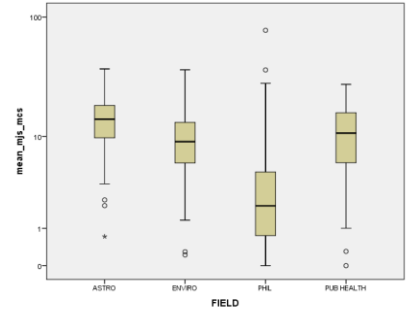 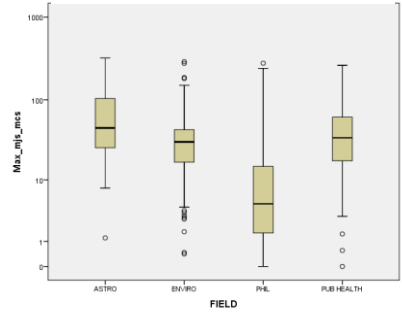 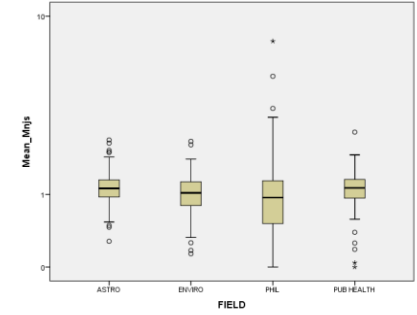

Summary of boxplots: minimum and maximum indicator scores

| Field | P | Fp | APP | Mean pp collab | Mean pp int collab | C | Csc | sc | %sc | nnc | %nc |
|---|---|---|---|---|---|---|---|---|---|---|---|
| **Astronomy** | 2-327 | 0.5-136.9 | 1.1-9.6 | 0-1 | 0-1 | 3-16481 | 1-12605 | 1-6189 | 9.7-77.5 | 0-36 | 0-66.6 |
| **Enviro. Sci** | 1-425 | 0.5-136.2 | 1-6.8 | 0-1 | 0-1 | 0-14141 | 0-11750 | 0-2391 | 0-87.5 | 0-33 | 0-100 |
| **Philosophy** | 1-140 | 0.3-140 | 1-4.8 | 0-1 | 0-1 | 0-3495 | 0-2934 | 0-809 | 0-100 | 0-114 | 0-100 |
| **Pub. Health** | 1-661 | 0.5-154.6 | 1-8.0 | 0-1 | 0-1 | 0-13520 | 0-11030 | 0-2490 | 0-66.6 | 0-173 | 0-100 |

| Field | CPP | SIG | Cless5 | PI | Cage | FC | FracCPP | Sum pptop n cits | h | AWCR | AWCRpa |
|---|---|---|---|---|---|---|---|---|---|---|---|
| **Astronomy** | 1-59.0 | 2-1365 | 0-10704 | 0-100 | 0-7.5 | 1.3-6734 | 0-64.6 | 0-192 | 1-66 | 0.7-2219.9 | 0.3-497.8 |
| **Enviro. Sci** | 0-51.1 | 0-1378 | 0-6958 | 0-121.7 | 0-8.1 | 0-4494.5 | 0-29.5 | 0-235 | 0-59 | 0-1425.1 | 0-400.7 |
| **Philosophy** | 0-33.6 | 0-360 | 0-1575 | 0-100 | 0-13 | 0-1895.6 | 0-59.3 | 0-65 | 0-32 | 0-354.5 | 0-163.3 |
| **Pub. Health** | 0-74.0 | 0-2040 | 0-8230 | 0-100 | 0.5.4 | 0-2797.4 | 0-63.1 | 0-243 | 0-60 | 0-1882.2 | 0-423.1 |

| Field | g | H2 | PopH | mquot | mgquot | Q2 | AW | e | m | A | AR | hg |
|---|---|---|---|---|---|---|---|---|---|---|---|---|
| **Astronomy** | 1-119 | 1.4-25.4 | 0-33 | 0.1-5 | 0.2-7.4 | 1.4-95.7 | 0.8-47.1 | 1-79.2 | 2-150 | 0-210.5 | 1.4-14.5 | 1-88.6 |
| **Enviro. Sci** | 0-99 | 0-24.1 | 0-27 | 0-1.8 | 0-3.0 | 0-82.3 | 0-37.7 | 0-59.6 | 0-1.8 | 0-149.7 | 0-12.2 | 0-76.4 |
| **Philosophy** | 0-56 | 0-15.1 | 0-17 | 0-1.6 | 0-2.4 | 0-42.8 | 0-18.8 | 0-39.2 | 0-1.6 | 0-94 | 0-9.6 | 0-42.3 |
| **Pub.Health** | 0-93 | 0-23.8 | 0-19 | 0-2.8 | 0-4.6 | 0-78.8 | 0-43.3 | 0-67.0 | 0-2.8 | 0-199.1 | 0-14.1 | 0-74.6 |

| Field | hnorm | ℏ | Sum pp top prop | NprodP | T>ca | mcs | mncs | Mean mjsmcs | Max mjsmcs | Mean mnjs |
|---|---|---|---|---|---|---|---|---|---|---|
| **Astronomy** | 0-7 | 1.2-90.7 | 0-80.0 | 0.2-260.1 | 0.2-22.4 | 0.3-58.2 | 0-4.7 | 0.7-37.5 | 1.2-320.5 | 0.2-2.3 |
| **Enviro. Sci** | 0-4 | 0-84.0 | 0-115.4 | 0-580.6 | 0-5.6 | 0-50.6 | 0-4.1 | 0.2-39.9 | 0.4-289.8 | 0.1-2.3 |
| **Philosophy** | 0-2 | 0-41.8 | 0-30.3 | 0-2718 | 0-4.3 | 0-53.4 | 0-25.1 | 0-77-9 | 0-277.3 | 0-7.6 |
| **Pub. Health** | 0-3 | 0-82.2 | 0-100.8 | 0-834.2 | 0-6.3 | 0-98.9 | 0-4.1 | 0-27.9 | 0-261.5 | 0-2.6 |